\documentclass[a4paper,12pt]{article}

\pdfoutput=1

\usepackage{amsmath}
\usepackage{amssymb}
\usepackage{hyperref}
\usepackage{graphicx}
\usepackage{cite}
\usepackage{color}

\makeatletter
\@addtoreset{equation}{section}
\renewcommand{\theequation}{\thesection.\@arabic\c@equation}
\makeatother

\begin{document}

\begin{titlepage}

\vspace*{-15mm}   
\baselineskip 10pt   
\begin{flushright}   
\begin{tabular}{r}    
{\tt OU-HET-1082}   
\end{tabular}   
\end{flushright}   
\baselineskip 24pt   
\vglue 10mm   

\begin{center}
{\Large\bf
Islands and stretched horizon
}

\vspace{8mm}   

\baselineskip 18pt   

\renewcommand{\thefootnote}{\fnsymbol{footnote}}

Yoshinori Matsuo$^{a,b,}$\footnote[2]{ymatsuo@gauge.scphys.kyoto-u.ac.jp}, 

\renewcommand{\thefootnote}{\arabic{footnote}}
 
\vspace{5mm}   

{\it  
$^a$ 
Department of Physics, Osaka University,\\1-1 Machikaneyama, Toyonaka, Osaka 560-0043, Japan
\\
$^b$ 
Department of Physics, Kyoto University,\\Kitashirakawa, Kyoto 606-8502, Japan
}
  
\vspace{10mm}   

\end{center}

\begin{abstract}
Recently it was proposed that the entanglement entropy of 
the Hawking radiation contains the information of a region including 
the interior of the event horizon, which is called ``island.'' 
In studies of the entanglement entropy of the Hawking radiation, 
the total system in the black hole geometry is separated into 
the Hawking radiation and black hole. 
In this paper, we study the entanglement entropy of the black hole 
in the asymptotically flat Schwarzschild spacetime. 
Consistency with the island rule for the Hawking radiation implies that 
the information of the black hole 
is located in a different region than the island. 
We found an instability of the island in the calculation of the entanglement entropy 
of the region outside a surface near the horizon. 
This implies that the region contains all the information of the total system 
and the information of the black hole is localized on the surface. 
Thus the surface would be interpreted as the stretched horizon. 
This structure also resembles black holes in the AdS spacetime with an auxiliary flat spacetime, 
where the information of the black hole is localized 
at the interface between the AdS spacetime and the flat spacetime. 
\end{abstract}

\baselineskip 18pt   

\end{titlepage}

\newpage

\baselineskip 18pt

\tableofcontents


\section{Introduction and summary}\label{sec:intro}

\begin{figure}
\begin{center}
\includegraphics[scale=0.3]{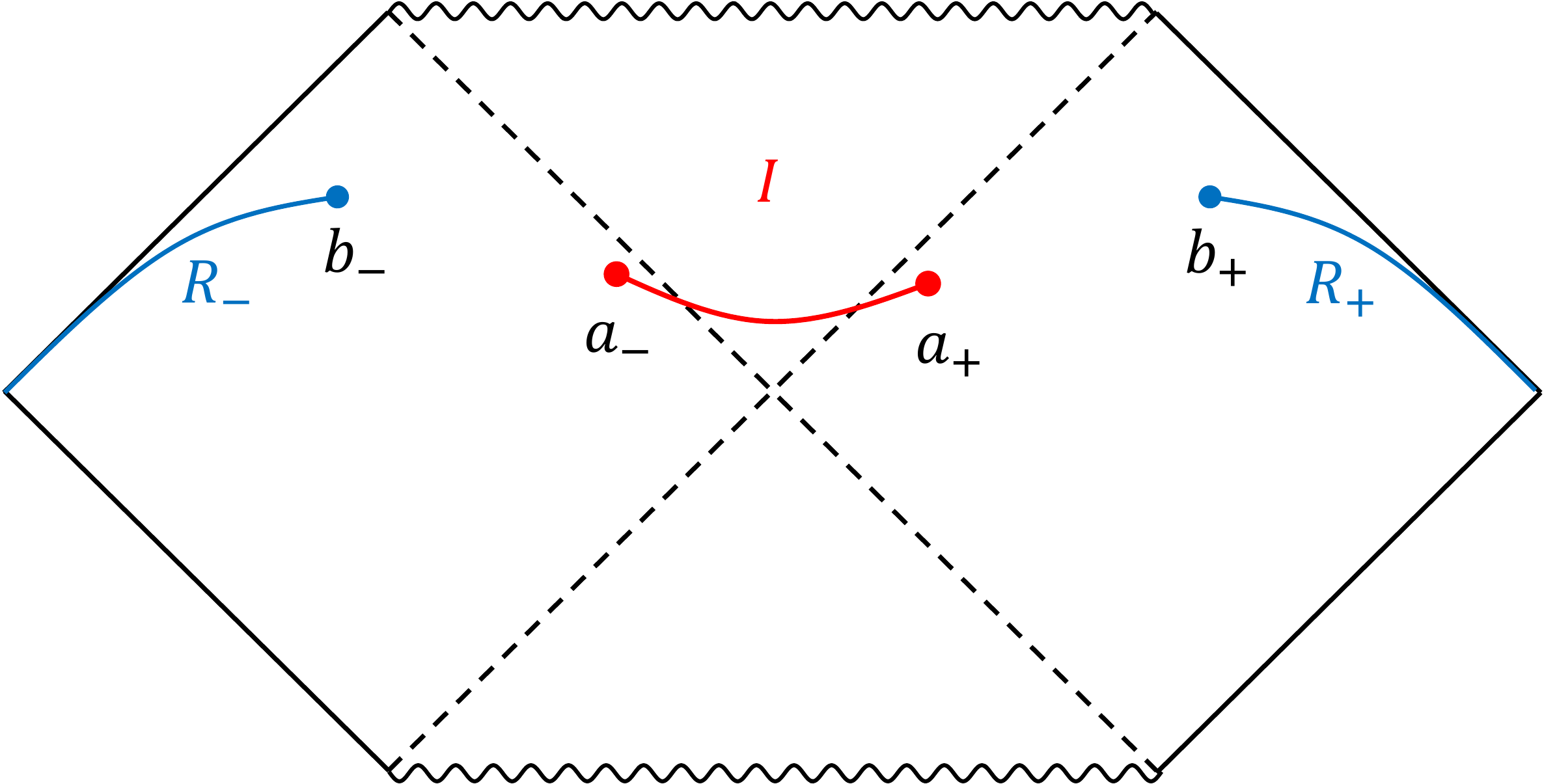}
\hspace{24pt}
\includegraphics[scale=0.3]{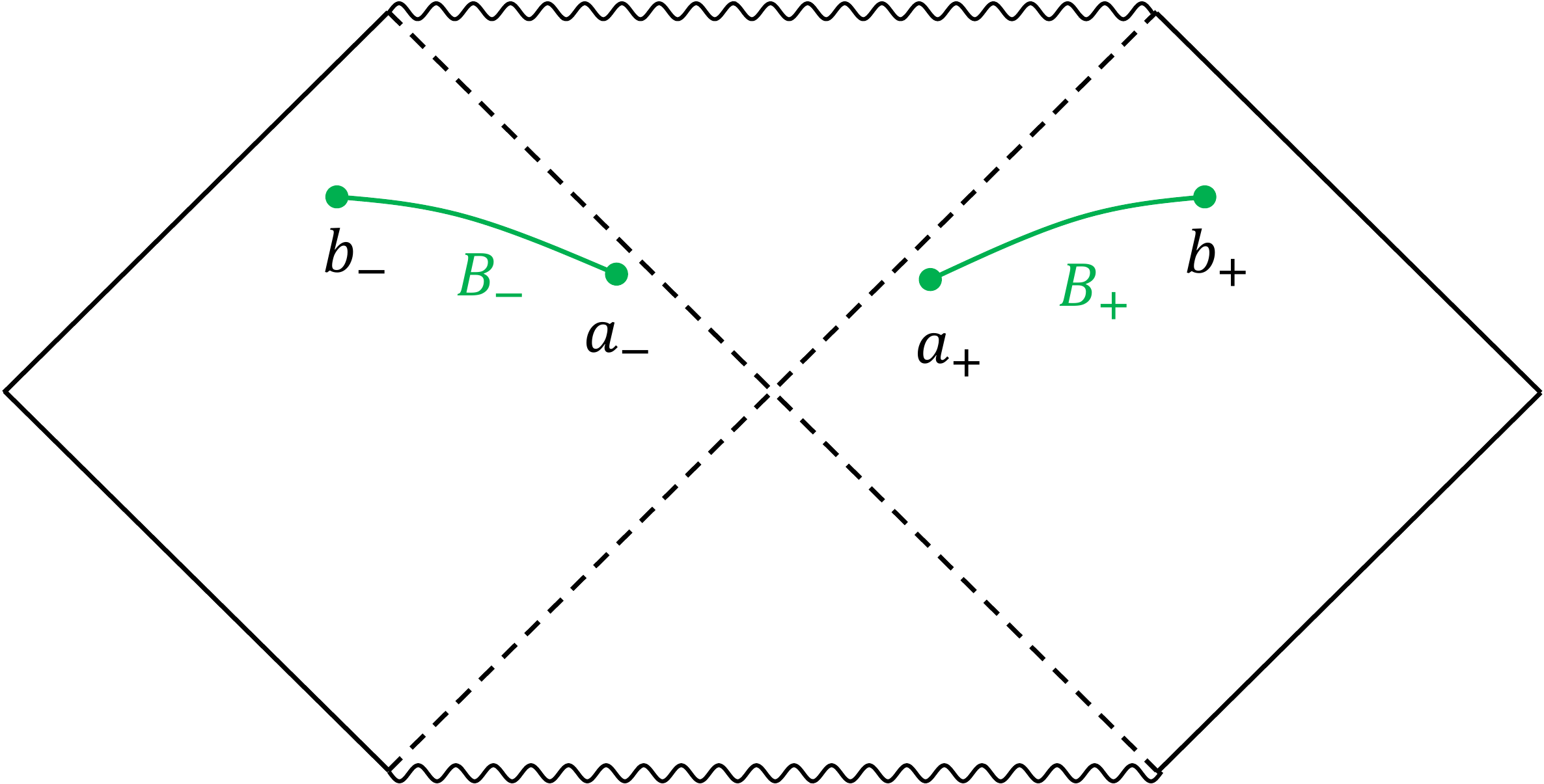}
\caption{%
The entanglement entropy of the Hawking radiation 
is identified with that in the region $R = R_+\cup R_-$ (left). 
After the Page time, the island $I$ appears as 
a consequence of the replica trick in gravitational theories. 
The region of the black hole $B = B_+\cup B_-$ is naively the complement of $R\cup I$ (right), 
which includes a region which is added by a similar mechanism to that for the island 
(see Fig.~\ref{fig:BH-crit}(right)). 
The boundaries between $R$ and $B$, $b_\pm$ are introduced by hand, 
while those of the island, $a_\pm$ are determined by 
the prescription of the quantum extremal surface. 
}\label{fig:BH}
\end{center}
\end{figure}

The information loss paradox is one of the most important problems in black hole physics 
\cite{Hawking:1976ra,Hawking:1974sw}. 
A recent progress in this field 
\cite{Penington:2019npb,Almheiri:2019psf,Almheiri:2019hni,Almheiri:2019yqk,Penington:2019kki,Almheiri:2019qdq} 
indicates that the Hawking radiation contains the information inside the event horizon.%
\footnote{
See \cite{Almheiri:2020cfm} for a review and 
\cite{Akers:2019nfi,Chen:2019uhq,Almheiri:2019psy,Chen:2019iro,Akers:2019lzs,Liu:2020gnp,Marolf:2020xie,Balasubramanian:2020hfs,
Bhattacharya:2020ymw,Verlinde:2020upt,Chen:2020wiq,Gautason:2020tmk,Anegawa:2020ezn,Hashimoto:2020cas,
Sully:2020pza,Hartman:2020swn,Hollowood:2020cou,Krishnan:2020oun,Alishahiha:2020qza,Banks:2020zrt,Geng:2020qvw,
Chen:2020uac,Chandrasekaran:2020qtn,Li:2020ceg,Bak:2020enw,Bousso:2020kmy,Anous:2020lka,Dong:2020uxp,
Krishnan:2020fer,Hollowood:2020kvk,Engelhardt:2020qpv,Karlsson:2020uga,Chen:2020jvn,Chen:2020tes,Hartman:2020khs,
Liu:2020jsv,Murdia:2020iac,Akers:2020pmf,Balasubramanian:2020xqf,Balasubramanian:2020coy,Sybesma:2020fxg,
Stanford:2020wkf,Chen:2020hmv,Ling:2020laa,Marolf:2020rpm,Harlow:2020bee,Akal:2020ujg,Hernandez:2020nem
} for related works. 
} 
The entanglement entropy of the Hawking radiation is identified with 
that of a region $R$ outside the black hole (See Fig.~\ref{fig:BH}), 
and then, it can be calculated by using the replica trick 
\cite{Callan:1994py,Holzhey:1994we,Calabrese:2009qy}. 
In theories with gravity, the most dominant saddle point possibly 
has a wormhole between different sheets of replica geometries, 
which gives the same effect to inserting a branch cut in addition to that on $R$. 
In this case, the entanglement entropy of $R$ effectively includes 
contributions from other regions, which are called islands $I$. 
The positions of the islands are determined such that 
the configuration becomes the saddle point. 
In the case of Einstein gravity, 
the boundaries of the islands are given by the quantum extremal surfaces \cite{Engelhardt:2014gca}. 
The entanglement entropy of the Hawking radiation is given by 
\begin{equation}
 S(R) 
 = 
 \min \left\{\mathrm{ext}\left[
 \frac{\text{Area}(\partial I)}{4G_N} 
 + S_\text{matter}(R\cup I)
 \right]\right\} \ . 
\end{equation}
Although the prescription of the (quantum) extremal surface was 
first proposed in the framework of holography \cite{Ryu:2006bv,Hubeny:2007xt}, 
the island rule would be applicable to black holes in more general theories. 

In this paper, we focus on the information of the black hole as the complement of the Hawking radiation, 
and show that the information of the black hole is localized on a surface, 
which is interpreted as the stretched horizon. 
We separate the total system into two subsystems: 
the Hawking radiation and the black hole (See Fig.~\ref{fig:BH}). 
The entanglement entropy of the Hawking radiation 
is identified with that of the region $R$. 
The quantum extremal surface becomes unstable 
when the region $R$ reaches a surface near the horizon. 
The instability implies that the region $R$ becomes continuous with the island $I$. 
Then, the entanglement entropy of the region $R$ is no longer 
that of the Hawking radiation but is identified with that of the total system. 
Therefore, the information of the black hole is localized on the surface, 
which would be identified with the stretched horizon. 


Since the entanglement entropy of the Hawking radiation is 
identified with the entanglement entropy of the region $R$, 
the entanglement entropy of the black hole is naively given by the entanglement entropy of the complement of $R$. 
However, studies in the case of black holes in the anti-de Sitter (AdS) spacetime 
implies that the entanglement entropy of the black hole 
is identified with that of a much smaller region. 

\begin{figure}
\begin{center}
\includegraphics[scale=0.3]{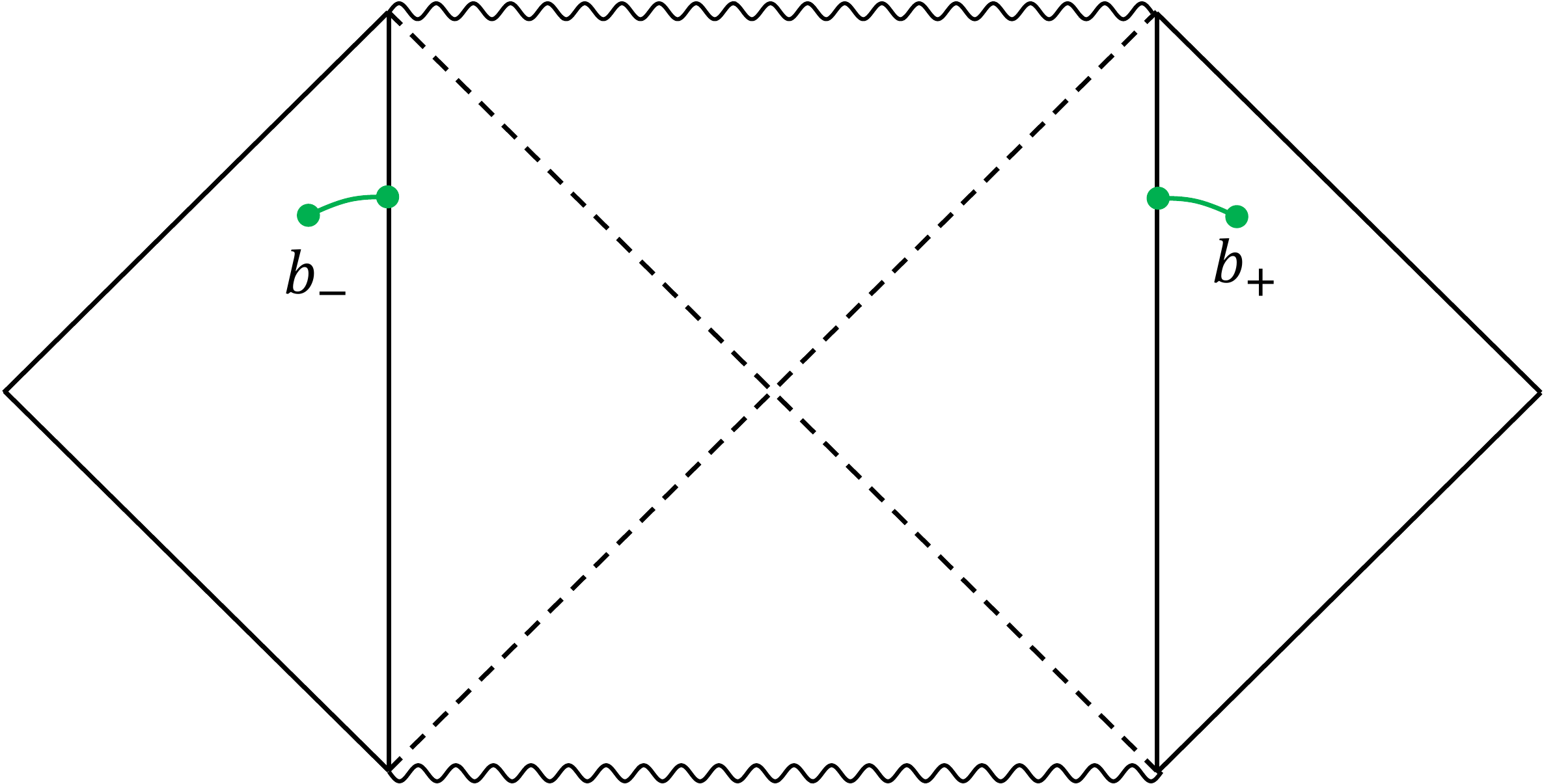}
\hspace{24pt}
\includegraphics[scale=0.3]{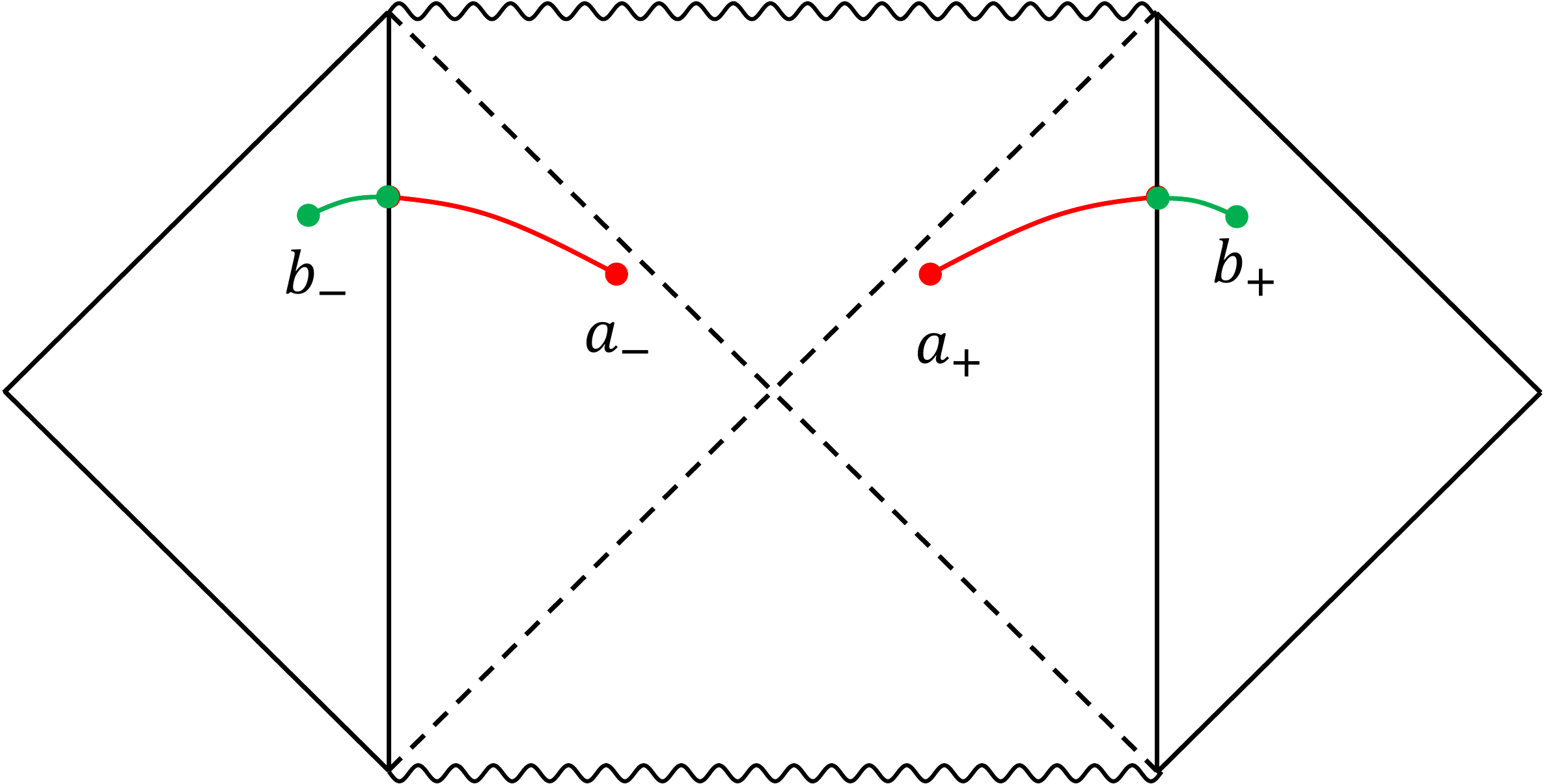}
\caption{%
In the case of black holes in AdS spacetime, 
an auxiliary system is introduced outside the boundary of AdS spacetime. 
The region of the black hole is put in the auxiliary flat spacetime (left). 
It is extended into the AdS spacetime due to gravitational effects, 
and ends at the quantum extremal surface after the Page time (right). 
Before the Page time, it continues to the region in the other side of the event horizon. 
For the entanglement entropy of a region only in one of the two auxiliary flat spacetime, 
the region in the AdS spacetime always ends at the extremal surface. 
}\label{fig:AdSBH}
\end{center}
\end{figure}

In the case of AdS spacetime, 
the Hawking radiation reaches the AdS boundary within a finite period of time. 
In order to study the Hawking radiation after it got out of the AdS spacetime, 
an interface to an auxiliary system on the flat spacetime is introduced at the AdS boundary 
\cite{Almheiri:2019psf,Almheiri:2019hni,Almheiri:2019yqk,Almheiri:2019qdq} (Fig.~\ref{fig:AdSBH}). 
The region $R$ of the Hawking radiation 
is defined in the auxiliary flat spacetime. 
The region of the black hole should 
be identified with the complement of $R$ in the flat spacetime in the auxiliary flat spacetime 
(Fig.~\ref{fig:AdSBH}(left)). 
When the entanglement entropy is calculated by using the replica trick, 
the branch cut is introduced only in the flat spacetime. 
The branch cut is automatically extended into the AdS spacetime by gravitational effects, 
since the gravitational solution in the AdS side is chosen such that it satisfies 
the boundary condition at the interface (Fig.~\ref{fig:AdSBH}(right)).%
\footnote{%
Although the effective region of the black hole $B$ is extended to the AdS spacetime, 
we are calculating the entanglement entropy of the region in the auxiliary flat spacetime. 
} 
In this case, the interface, or equivalently, the boundary of the AdS spacetime 
has the information of the black hole because of the AdS/CFT correspondence. 

\begin{figure}
\begin{center}
\includegraphics[scale=0.3]{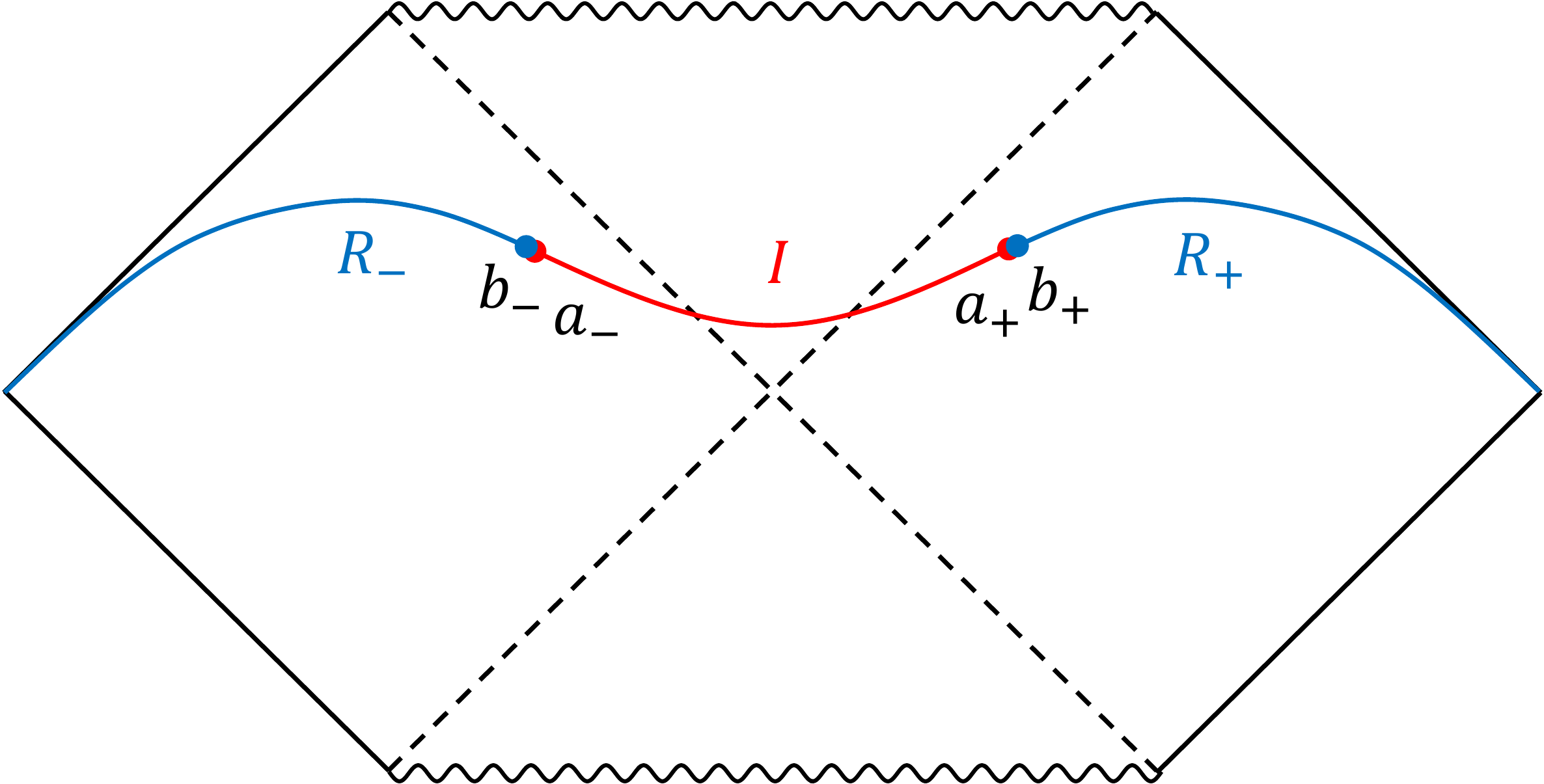}
\hspace{24pt}
\includegraphics[scale=0.3]{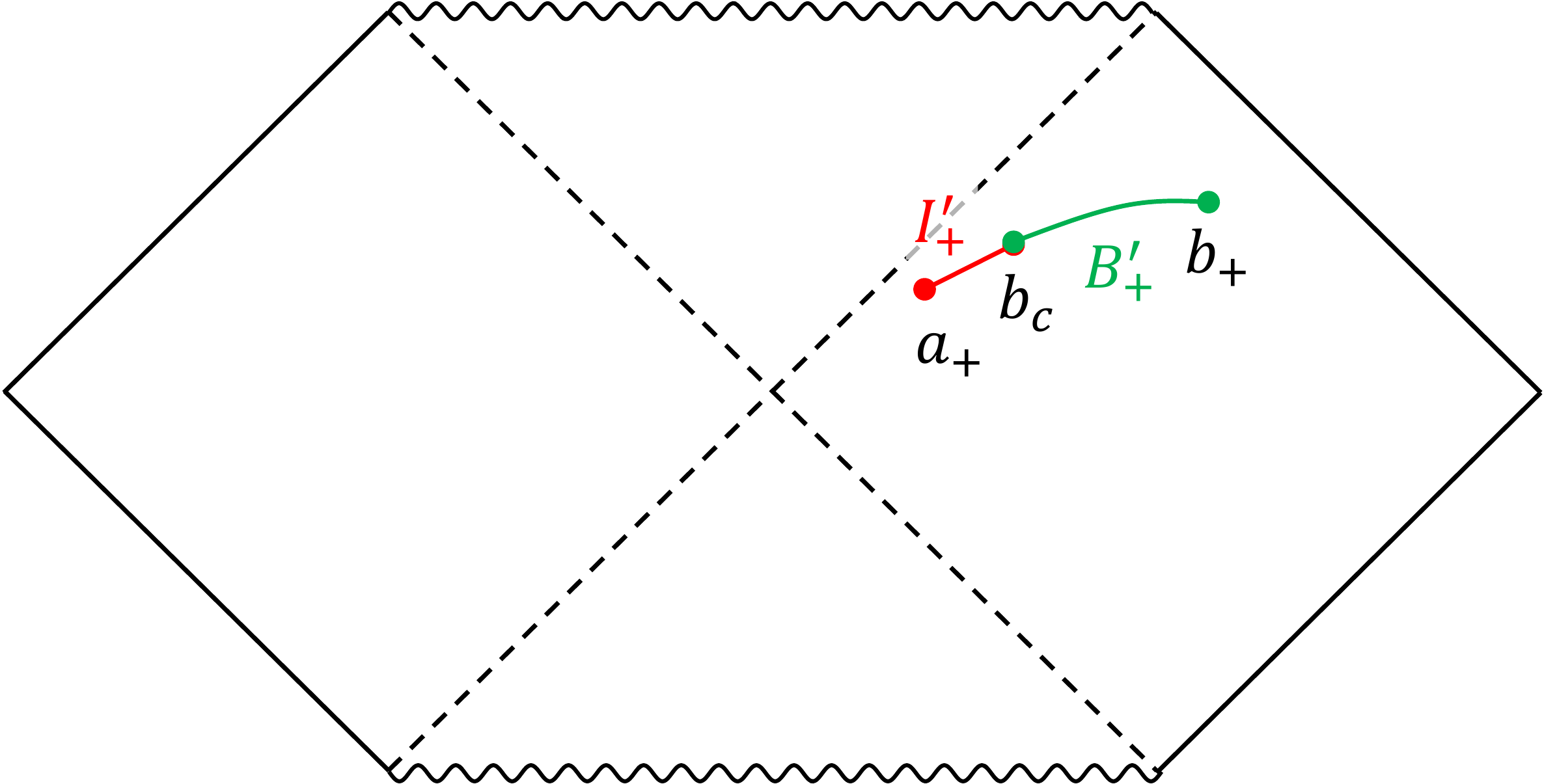}
\caption{%
The island can be maximally extended by putting $b_\pm$ as close to $a_\pm$ as possible (left). 
The entanglement entropy becomes zero when $b_\pm$ is identical to $a_\pm$, 
while the quantum extremal surface becomes unstable 
when the radius at $b_\pm$ becomes $b_c$, implying that 
the quantum extremal surface rolls down to a global minimum at $b_\pm$. 
A part of $B$ overlaps with the maximally extended island. 
This part $I'$ in $B$ is interpreted as a consequence of the replica trick, 
and referred to as ``hidden island'' in this paper. 
Thus, the region $B$ consists of 
the essential region of the black hole $B'$ and the hidden island $I'$ (right). 
The information of the black hole is localized at the stretched horizon at $b_c$. 
}\label{fig:BH-crit}
\end{center}
\end{figure}

The entanglement entropy of the black hole in the asymptotically flat Schwarzschild spacetime 
is naively identified with that of $\overline R$, the complement of the region $R$. 
As in the case of AdS, the region $I$ in $\overline R = B\cup I$ can be interpreted as 
the island in the sense that it appears as a consequence of the replica trick in the gravitational theory. 
Furthermore, the region $B$ still contains ``hidden islands'' $I'$, 
which is also involved by gravitational effects on the replica geometries (Fig.~\ref{fig:BH-crit}(right)), 
as a part of $B$ in AdS appears due to the gravitation in the AdS side. 
Thus, the entanglement entropy of the black hole is identified with that of the region $B'$, 
where $B = B'\cup I'$. 
The region $B'$ is effectively extended to $B\cup I$ before the Page time and to $B$ after the Page time, 
when we use the replica trick and take effects of gravity into account. 

In order to see that $I$ in $\overline R = B\cup I$ in the black hole before the Page time 
can be interpreted as the island, or equivalently, a consequence of the replica trick, 
we separate the black hole into that in the right wedge and that in the left wedge.%
\footnote{%
The eternal black hole geometry is two-sided --- 
it has two exteriors of the event horizon. 
Here, we call them as right wedge and left wedge. 
The Hawking radiation can be further separated into two subsystems: 
that in the right wedge and that in the left wedge, 
and they are identified with the $R_+$ and $R_-$, respectively. 
Here, we assume that the black hole can also be separated into two subsystems $B_+$ and $B_-$. 
The subscript ``$+$'' (``$-$'') stands for the subregion in the right (left) wedge. 
}
After the Page time, the island $I$ appears in the entanglement entropy of the Hawking radiation $R$. 
However, if we consider the Hawking radiation in the right wedge only, 
no island appears even after the Page time. 
Similarly, before the Page time, the entanglement entropy of 
the black hole is given by that of $\overline R = B \cup I$, 
but if we consider the black hole in the right wedge only, 
the region $I$ does not appear and the entanglement entropy 
is given by that of $B_+$ even before the Page time. 
This implies that the region $I$ in $\overline R = B\cup I$ appears as a consequence of the replica trick, 
and hence, can be interpreted as the island,%
\footnote{%
In this paper, the term ``island'' does not refer only to the region $I$, 
but stands for regions which are involved by gravitational effects in the replica trick. 
}
also in the calculation of the entanglement entropy of the black hole before the Page time. 

The region $B_\pm$ still contains the hidden island $I'_\pm$, 
which is involved as a consequence of the replica trick. 
The inner boundary of the region $B_\pm$ is given by the quantum extremal surface. 
This implies that the region of the black hole is originally a smaller region $B'$, 
but is extended 
by gravitational effects in the replica geometries. 
%
The extended part is the hidden island $I'_\pm$, and hence, 
the end point of $I'_\pm$ is given by the quantum extremal surface. 
Thus the region $B'_\pm$ effectively becomes $B_\pm = B'_\pm\cup I'_\pm$ (Fig.~\ref{fig:BH-crit}). 
In the case of AdS, the state of the black hole subsystem can be specified by 
the information in the auxiliary flat spacetime and the interface, which correspond to $B'_\pm$. 
In a similar fashion, the black hole state in the case of asymptotically flat spacetime 
can be identified by the information only in $B'=B'_+\cup B'_-$. 

In order to specify the hidden islands $I'_\pm$ in $B_\pm$, 
we consider the limit in which the region of the black hole $B$ disappears. 
We take the boundary $b_\pm$ between $R$ and $B$ to inner places in the spacetime, 
and then, the quantum extremal surface $a_\pm$ moves to outer positions (Fig.~\ref{fig:BH-crit}(left)). 
This corresponds to identifying more degrees of freedom 
in the total system as that of the Hawking radiation. 
It is expected that $b_\pm$ eventually coincides with $a_\pm$ at some point, 
and then, $R$ merges to $I$, or equivalently, $B$ disappears. 
The region $R$ is identified with the total system and the entanglement entropy is zero. 
In the case of AdS, this happens when we take $b_\pm$ to the interface at the AdS boundary. 
The surface $b_\pm$ for $a=b$ corresponds to the interface in the case of AdS. 

There is a subtlety when the region $R$ merges with the island $I$.  
We find that the quantum extremal surface becomes unstable, 
when $a_\pm$ and $b_\pm$ are sufficiently close to each other and the radius at $b_\pm$ 
takes some critical value $b_c$. 
Because of this instability, the quantum extremal surface $a$ jumps to $a=b$ for $b=b_c$, 
and then, the entanglement entropy also jumps to zero, discontinuously. 
If $b$ is slightly larger than $b_c$, the entanglement entropy of $B_\pm$ 
still approximately equals to the Bekenstein-Hawking entropy. 
This implies that the information of the black hole is localized 
on the surface at $r = b_c$, which can be interpreted as the stretched horizon. 
For $b>b_c$ the effective region of the black hole $B$ extends in $a<r<b$, 
but the state of the black hole can be specified only by the information in $B'$, 
or equivalently, $b_c \leq r \leq b$, and 
the other region, $r < b_c$ in $B$ is identified as the hidden island $I'$ (Fig.~\ref{fig:BH-crit}(right)). 

This paper is organized as follows. 
In Sec.~\ref{sec:review}, we review on the island in the calculation of 
the entanglement entropy of the Hawking radiation 
in the asymptotically flat four-dimensional Schwarzschild spacetime. 
In Sec.~\ref{sec:region}, we divide the Hawking radiation and that of the black hole 
further into those in the right and left wedges in the two-sided spacetime, 
namely, $R_+$, $R_-$, $B_+$ and $B_-$. 
We study the entanglement entropy of each of 
the four regions $R_+$, $R_-$, $B_+$ and $B_-$, and that of their unions. 
Although the effective region of a union of two regions is different from 
the union of the two effective regions since the associated region of islands are 
different for different combination of the regions, 
the entanglement entropy satisfies the consistency conditions: 
subadditivity condition and strong subadditivity condition. 
In Sec.~\ref{sec:stretch}, we consider the island in $B$ and 
show that the quantum extremal surface has instability at $b=b_c$. 
Sec.~\ref{sec:discussion} is devoted for discussions. 
We also comment on an analogous structure of regions and islands 
to the entanglement wedge in the bulk of holography in Appendix~\ref{sec:analogy}. 
In Appendix~\ref{sec:position}, we discuss about the position of the island in the case of 
the black hole which is formed by the gravitational collapse.


\section{Islands in Schwarzschild geometry}\label{sec:review}

Here, we first review on the island in the Schwarzschild black holes \cite{Hashimoto:2020cas}. 
In this paper, we consider the islands in the eternal Schwarzschild geometry. 
The gravitational action is given by the Einstein-Hilbert action 
with the Gibbons-Hawking term; 
\begin{align}
 I &= I_\text{gravity} + I_\text{matter} \, ,
 \\
 &I_\text{gravity} 
 = 
 \frac{1}{16\pi G_N} \int_{\mathcal M} d^4 x \sqrt{-g} \, R 
 + \frac{1}{8\pi G_N} \int_{\partial \mathcal M} d^3 x \sqrt{-h} \, K \ , 
\end{align}
where $G_N$ is the Newton constant. 
The Schwarzschild metric is given by 
\begin{equation}
 ds^2 
 = 
 - \left(1 - \frac{r_h}{r}\right) dt^2 
 + \left(1 - \frac{r_h}{r}\right)^{-1} dr^2
 + r^2 d \Omega^2 \ , 
\end{equation}
where $r_h$ is the Schwarzschild radius. 
The Hawking temperature $T_H$ is given by $\beta = 1/T_H = 4\pi r_h$. 
It is convenient to use the Kruskal coordinates; 
\begin{equation}
 ds^2 = - \frac{dU dV}{W^2} + r^2 d \Omega^2 \ , 
\end{equation}
where
\begin{align}
 W = \sqrt{\frac{r}{4r_h}\frac{U V}{r-r_h}} = \sqrt{\frac{r}{4r_h^3}} \, e^{\frac{r-r_h}{2r_h}} \ , 
\end{align}
and the coordinates are defined as 
\begin{align}
 r_* 
 &= 
 r- r_h + r_h \log \frac{r-r_h}{r_h} \ , 
\\
 U 
 &= 
 - e^{-\frac{t - r_*}{2r_h}} 
 = - \sqrt{\frac{r-r_h}{r_h}} e^{-\frac{t-(r-r_h)}{2r_h}} \ , 
&
 V 
 &= 
 e^{\frac{t + r_*}{2r_h}} 
 = \sqrt{\frac{r-r_h}{r_h}} e^{\frac{t+(r-r_h)}{2r_h}} \, .
\end{align}

The entanglement entropy of a region $R$ can be calculated by using the replica trick. 
When the replica trick is applied for the gravitational theories, 
configurations with wormholes which connect 
different copies of the spacetime should be taken into account. 
The entanglement entropy of the region $R$ for configurations with such replica wormholes 
is equivalent to the entanglement entropy of the region $R\cup I$ 
for the configurations without replica wormholes. 
The region called island, $I$, is identified to the region where the wormhole is located. 
Thus, we effectively calculate the entanglement entropy of $R\cup I$ 
when the replica wormhole configuration dominates in the path integral, 
although we really consider the entanglement entropy of the region $R$. 
Since the maximal partition function for the replica geometries gives 
the minimal entanglement entropy, the configuration 
which gives minimum of the entropy dominates in the path integral. 
The positions (of the boundaries) of the islands should be chosen 
to extremize the entanglement entropy, and then, 
the minimum of these extrema gives the leading order contribution. 

At the leading order of the small $G_N$ expansion, 
the replica trick for the gravity part simply gives the area term, 
which is proportional to the area of the boundary of the regions 
\cite{Faulkner:2013ana,Lewkowycz:2013nqa,Dong:2016hjy,Dong:2017xht}. 
The entanglement entropy of the matter part consists of 
UV-divergent local terms and finite non-local terms \cite{Bombelli:1986rw,Srednicki:1993im}. 
The leading contribution of the local terms takes the same form 
to the area term of the gravity part, and can be absorbed by 
the renormalization of the Newton constant \cite{Susskind:1994sm}. 
The non-local terms come from the effects of the correlations 
between the twist operators at the boundaries of the regions. 
Thus, the entanglement entropy is expressed as 
\begin{equation}
 S
 = 
 \sum_{\partial R,\ \partial I} \frac{\text{Area}}{4 G_N} 
 + \sum_{\partial R,\ \partial I} S_\text{matter}^\text{(non-local)} \ , 
\end{equation}
where the summation should be taken over all the boundaries of the region $R$ 
and, if the configuration with the islands dominates, those of the islands $I$. 
When the distance between a pair of twist operators is short, 
the non-local term for that pair is approximated as \cite{Casini:2005zv,Casini:2009sr}
\begin{equation}
 S_\text{matter}^\text{(non-local)} = \pm \kappa \,c \, \frac{\text{Area}}{L^2} \ , 
 \label{S-short}
\end{equation}
where $L$ is the proper distance between the twist operators. 
The sign is plus for twist operators in the same orientation, 
and is minus for opposite twists. 
If the distance between the twist operators is long, 
we assume that the correlation can be approximated by 
that of the s-waves, 
and then, the non-local term is approximately expressed as 
\begin{equation}
  S_\text{matter}^\text{(non-local)} 
  = 
  \mp \frac{c}{6} \log \frac{\left(U(y) - U(x)\right)\left(V(x) - V(y)\right)}{W(x)W(y)} \ , 
  \label{S-long}
\end{equation}
where $x$ and $y$ indicates the positions of the pair of the twist operators. 
Here, the sign is plus for opposite twists. 

Now, we calculate the entanglement entropy of the region $R$, 
which is identified to that of the Hawking radiation (See Fig.~\ref{fig:BH}). 
The region $R$ is defined by $r\geq b$, 
and the time at the boundaries $b_+$ of the region $R$ in the right wedge is referred to as $t_b$. 
(The boundary in the left wedge, $b_-$, is located at $(t,r) = (- t_b + i \beta/2,b)$.) 
The area terms at $b_\pm$ are usually excluded 
from the entanglement entropy of the Hawking radiation 
since these terms are present even in the case of the empty flat spacetime. 
In this paper, we include them explicitly since 
they are necessary to see the (strong) subadditivity condition of the entanglement entropy. 

For the configuration without islands, the distance between two twist operators is long. 
The entanglement entropy is calculated by using the formula \eqref{S-long} as 
\begin{equation}
 S = \frac{2\pi b^2}{G_N} 
 + \frac{c}{6} \log \left[\frac{16r_h^2(b-r_h)}{b}\cosh^2\frac{t_b}{2 r_h}\right] \ . 
\end{equation}
For the configuration with an island, the positions of boundaries of the island $a_\pm$ 
are determined such that the entanglement entropy is extremized. 
Here the coordinates at $a_+$ are referred to as $(t,r) = (t_a, a)$, 
and then, the other boundary $a_-$ is located at $(t,r) = (- t_a + i \beta/2,a)$. 
For sufficiently large $t_b$, interactions between twist operators in the right wedge 
and those in the left wedge is negligible since 
the distance between the right and left wedge becomes very large.
For $b \gg r_h$, the entanglement entropy is calculated by using \eqref{S-long} 
and is extremized when $t_a = t_b$ and 
\begin{equation}
 a 
 \simeq 
 r_h 
 + \frac{(c\,G_N)^2}{144\pi^2 r_h^2 (b-r_h)} e^{\frac{r_h-b}{r_h}} \ . 
 \label{long-a}
\end{equation}
Then, the entanglement entropy is given by 
\begin{align}
 S 
 &\simeq 
 \frac{2\pi r_h^2}{G_N} + \frac{2\pi b^2}{G_N}
 + \frac{c}{6} 
 \left[
 \log \left(\frac{16 r_h^3 (b-r_h)^2}{b} \right) + \frac{b-r_h}{r_h}  
 \right] \ . 
\end{align}
If the boundaries of $R$ is close to the event horizon, 
we use \eqref{S-short} instead, and 
the boundaries of the island are located at 
\begin{align}
 a \simeq r_h + \frac{( \kappa c \, G_N)^2}{4(b-r_h)^3} \ . 
 \label{short-a}
\end{align}
Then, the entanglement entropy is calculated as 
\begin{align}
 S 
 \simeq 
 \frac{2\pi r_h^2}{G_N} + \frac{2\pi b^2}{G_N} 
 - 2\pi  \kappa \, c \, \frac{r_h}{b-r_h} \  .
\end{align}

In either case of $b\gg r_h$ or $b \gtrsim r_h$, 
the boundaries of the island are located at $a = r_h + \mathcal O(\ell_p^4/r_h^3)$, 
where $\ell_p = G_N^{1/2}$ is the Planck length,  
and the entanglement entropy for the configuration with the island 
is approximated at the leading order as 
\begin{equation}
 S = 2S_\text{BH} + \frac{2\pi b^2}{G_N} 
\end{equation}
where the Bekenstein-Hawking entropy is given by 
\begin{equation}
 S_\text{BH} = \frac{\pi r_h^2}{G_N} \ . 
\end{equation}
Since the entanglement entropy for the configuration without islands 
is approximated for sufficiently large $t_b$ as 
\begin{equation}
 S \simeq \frac{2\pi b^2}{G_N} + \frac{c}{6 r_h}\, t_b \ , 
\end{equation}
the configuration with the island dominates over that without islands 
after the Page time $t_p$ where 
\begin{equation}
 t_p = \frac{3 S_\text{BH}}{\pi c T_H} \ , 
\end{equation}
while the configuration without the islands dominates before the Page time. 
Thus, the entanglement entropy of the Hawking radiation 
follows the Page curve \cite{Page:1993wv,Page:2013dx}.


\section{Entanglement entropy of subsystems}\label{sec:region}

By using the island rule, the entanglement entropy of the Hawking radiation is 
identified with the entanglement entropy of the region $R$, 
but the region effectively becomes $R\cup I$ 
as the island $I$ is involved by gravitational effects in the replica trick.%
\footnote{%
We call the region including additional parts 
involved by gravitational effects as the ``effective region.'' 
Although the entanglement entropy of the Hawking radiation is identified with that of the regoin $R$, 
the formula of the island rule contains contributions from the island $I$. 
Hence, we call the region $R$ as the original region and 
$R\cup I$ as the effective region of the Hawking radiation. 
} 
Since the entanglement entropy of a subsystem must equal to that of its complement, 
the effective region which gives the entanglement entropy of the black hole%
\footnote{%
Here, the black hole does not mean the total system in the black hole geometry 
but the complement of the Hawking radiation. 
In the derivation of the Page curve \cite{Page:1993wv,Page:2013dx}, 
the total system is divided into two subsystems: the Hawking radiation and the black hole.
} 
must be the complement of the effective region of the Hawking radiation. 
Thus, the effective region of the black hole is the complement of $R$ before the Page time, 
but is the complement of $R\cup I$ after the Page time. 
In this section, we discuss that the region $I$ in the effective region 
of the black hole before the Page time can also be interpreted as the island, 
in the sense that it appears as a consequence of gravitational effects, 
also in the calculation of the entanglement entropy of the black hole. 

In the case of the AdS, 
the Hawking radiation reaches the AdS boundary within a finite period of time, and hence, 
we introduce an interface with the auxiliary flat spacetime at the AdS boundary. 
In this case, the region of the black hole is given by the complement of $R$ in the flat spacetime. 
Although the region is originally defined only in the flat spacetime, 
it is effectively extended into the AdS spacetime. 
When we calculate the entanglement entropy of that region by using the replica trick, 
we insert the branch cut there. 
The region includes the interface between the AdS spacetime and flat spacetime, 
and imposes a non-trivial boundary condition to the AdS side of the replica geometries. 
The configuration of the replica geometry in the AdS side is determined 
to satisfy the boundary condition, and hence, the branch cut is extended into the AdS spacetime. 
Since, the effective region inside the AdS spacetime is determined 
by the prescription of the quantum extremal surface, 
the effective region of the black hole is given by 
the complement of the effective region of the Hawking radiation. 

\begin{figure}
\begin{center}
\includegraphics[scale=0.3]{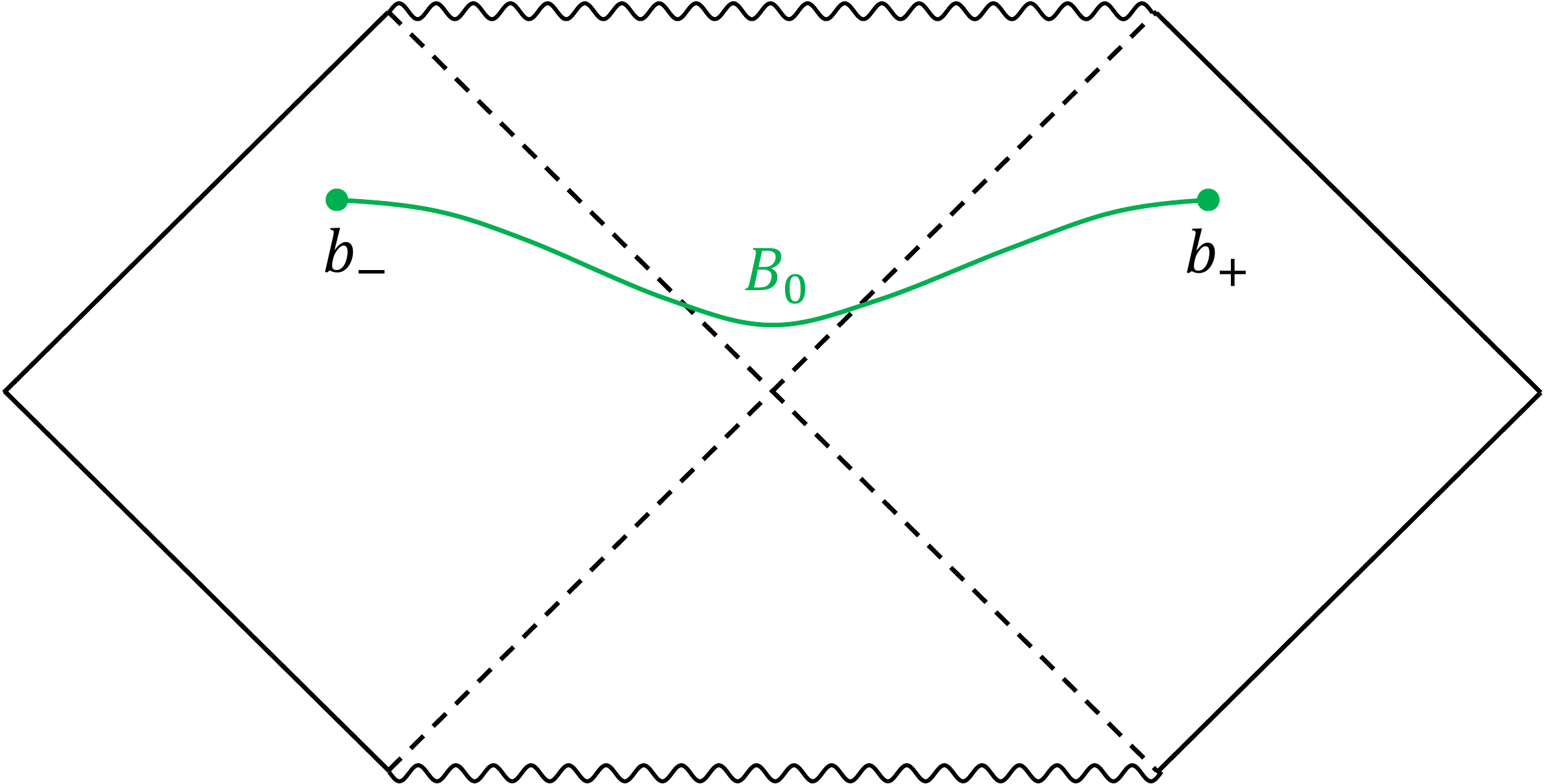}
\hspace{24pt}
\includegraphics[scale=0.3]{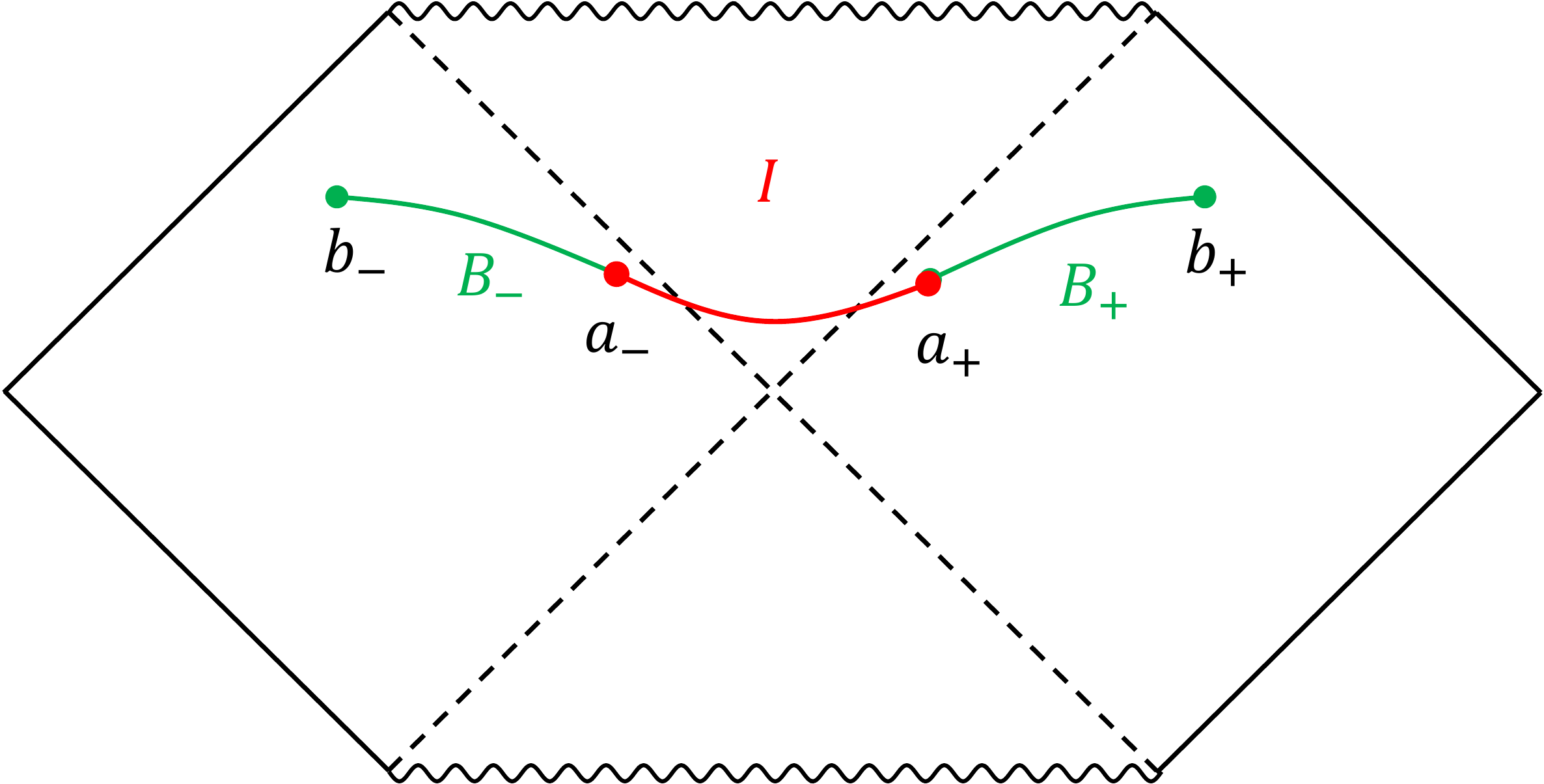}
\caption{%
The statement that the island appears after the Page time 
naively sounds as if it is absent before the Page time. 
Then, the region of the black hole subsystem $B$ covers all the other region than $R$, namely, $B = B_0$ (left). 
However, the black hole subsystem would be able to separated 
into that in the left wedge $B_-$ and that in the right wedge $B_+$. 
Information of the island $I$ cannot be reproduced by either of $B_+$ or $B_-$, 
and the region $B$ contains the island $I$, namely, $B_0 = B \cup I$, where $B = B_+ \cup B_-$ (right). 
}\label{fig:BH-B}
\end{center}
\end{figure}

In the case of the asymptotically flat Schwarzschild spacetime, 
the Hawking radiation does not reach the boundary within a finite period of time, 
and hence, the region should be defined within the Schwarzschild spacetime itself. 
However, the complement of the region $R$ in the Schwarzschild spacetime 
includes the island $I$, and hence, does not give 
a consistent entanglement entropy of the black hole after the Page time. 
Hence, we determine the effective region of the black hole 
as follows. 
We divide the system into two subsystems introducing the boundary $b_\pm$ 
between the region of the Hawking radiation and that of the black hole. 
If branch cuts on the replica geometries are introduced just inside the boundaries $b_+$ and $b_-$, 
they will be extended by the gravitational effect so that 
the configuration becomes a saddle point of the path integral. 
There are two possible configurations; 
the branch cuts are merged with each other to form a single branch cut, 
or, the branch cuts end at the quantum extremal surface in each sides of the event horizon. 
The former gives the dominant saddle point in the path integral before the Page time, 
and the latter is dominant after the Page time. 
%
This implies that the effective region of the black hole after the Page time, 
namely, the complement of $R\cup I$ would be more essential than that before the Page time.%
\footnote{%
Here, ``essential'' means that the region has the information. 
In the calculation of the entanglement entropy of the Hawking radiation, 
the entanglement entropy of $R$ is effectively given by $R\cup I$, 
implying that $R$ has the information of $I$. 
Then, $R$ is more essential since the information of $I$ can be read off from $R$, 
but the information of $R$ cannot be read off from $I$. 
} 
We refer to this region as $B$. 
The effective region before the Page time is given by $B\cup I$ but 
the region $I$ would be interpreted as the island, 
in the sense that $I$ appears as a consequence of the gravitational effects (Fig.~\ref{fig:BH-B}). 

In order to see that the island $I$ is involved by gravitational effects, 
we separate the region $R$ and $B$ into those in the right wedge and those in the left wedge, namely, 
$R$ into $R_+$ and $R_-$, and $B$ into $B_+$ and $B_-$, respectively. 
The island $I$ appears in the calculation of 
the entanglement entropy of $R = R_+ \cup R_-$ after the Page time, 
but no island appears the entanglement entropy of either $R_+$ or $R_-$. 
In a similar fashion, before the Page time, the entanglement entropy of $B = B_+\cup B_-$ 
is effectively given by the entanglement entropy of $B\cup I$. 
The entanglement entropy of either $B_+$ or $B_-$ does not involve the region $I$, even before the Page time. 
This is because $I$ in the entanglement entropy of $B$ before the Page time 
is involved by gravitational effects in the replica trick. 
We consider the effective regions of each of $R_+$, $R_-$, $B_+$ and $B_-$ 
and calculate the entanglement entropy. 
We also evaluate the entanglement entropy of unions of arbitrary combination 
of the subsystems above to check the subadditivity condition and strong subadditivity condition. 

\subsection{Entanglement entropy of single connected region}


\subsubsection*{Entanglement entropy of $R_\pm$}

\begin{figure}
\begin{center}
\includegraphics[scale=0.3]{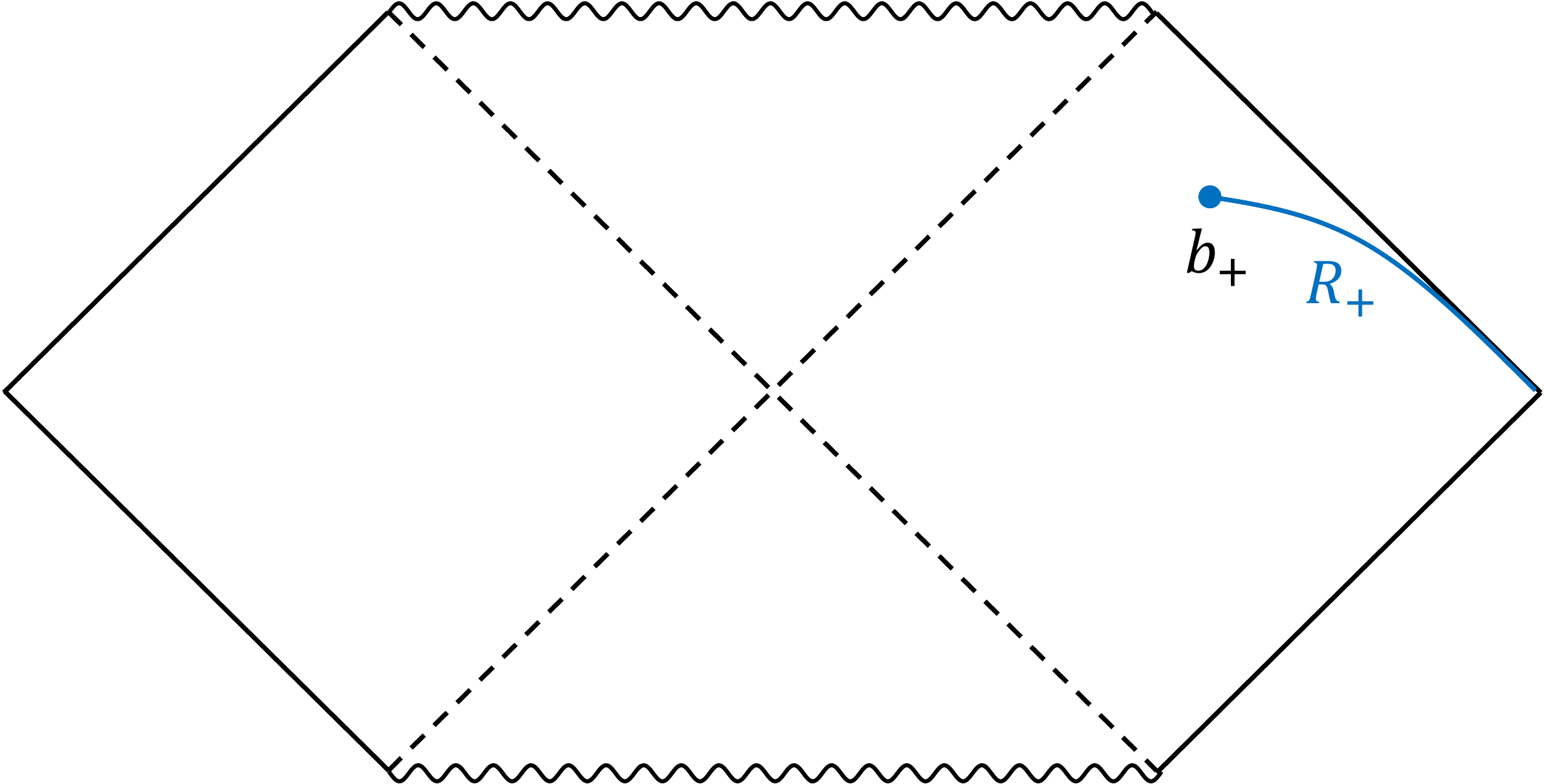}
\hspace{24pt}
\includegraphics[scale=0.3]{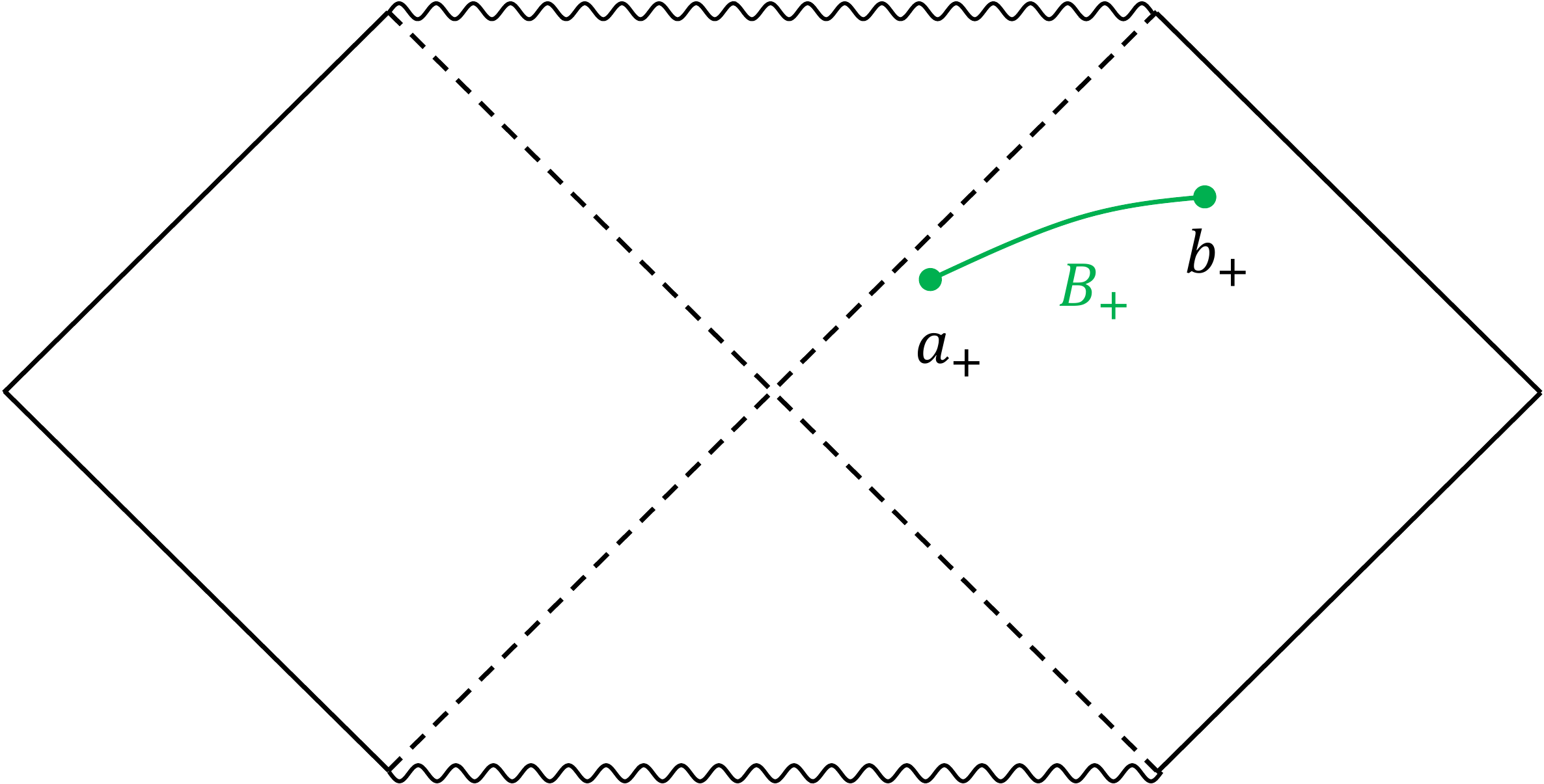}
\caption{%
The effective region of the Hawking radiation only in the right wedge $R_+$ (left)
and that of the black hole (right). 
The entanglement entropy of $R_+$ contains no contribution from the island even after the Page time. 
The black hole subsystem $B$ can also be separated into that in the right wedge $B_+$ and that in the left wedge $B_-$. 
The effective region of $B_+$ does not includes the island, implying that 
the black hole subsystem in each side has no information of the island. 
}\label{fig:BH-r}
\end{center}
\end{figure}

We calculate the entanglement entropy of the single region of $R_+$ or $R_-$ (Fig.~\ref{fig:BH-r}(left)). 
Since the regions $R_+$ and $R_-$ are related 
to each other by the exchange of $U$ and $V$, we focus on $R_+$. 
The region $R_+$ is defined by the inner boundary at $(t,r) = (t_b,b)$ 
and it extends to the spatial infinity, $r\to\infty$. 
The quantum extremal surface can appear only inside the inner boundary. 
However, for the island $I$ to be a finite region, 
there must be two quantum extremal surfaces for two boundaries of the island. 
It is straightforward to see that there is no stable configuration with two extremal surfaces. 
Assuming that the island cannot extend to the asymptotic infinity, 
no island can appear in the entanglement entropy of $R_+$. 

For $R_+$, there is only one twist operator. 
In order to calculate the one-point function of the twist operator, 
it is convenient to introduce an IR cut-off (boundary) of the spacetime 
and to consider the correlation with the mirror image on the other side of the IR cut-off. 
By putting the IR-cut off at $r=2 r_h\log \frac{\Lambda}{2}$, 
the entanglement entropy of $R_+$ (and that of $R_-$) is calculated as 
\begin{equation}
 S
 = 
 \frac{\pi b^2}{G_N} 
 + \frac{c}{6} \log \Lambda \ . 
\end{equation}
It is obvious that the entanglement entropy of $R_+$ is independent of time, 
since it depend only on the time in the right wedge, which can be absorbed by 
using the time translation invariance of the eternal Schwarzschild spacetime. 

By taking the effect of the IR cut-off, 
the entanglement entropy of $R_\pm$ satisfies 
the subadditivity condition with that of $R = R_+\cup R_-$. 
The IR cut-off must be larger than the distance between $b_+$ and $b_-$ at the Page time. 

\subsubsection*{Entanglement entropy of $B_\pm$}

The region $B_+$ is defined by two spherically symmetric boundaries, $a_+$ and $b_+$ (Fig.~\ref{fig:BH-r}(right)). 
The outer boundary $b_+$ is given by $(t,r) = (t_b, b)$, and 
the inner boundary $a_+$ is the quantum extremal surface, $(t,r) = (t_a,a)$. 
For configurations with the islands, three quantum extremal surfaces are necessary. 
Since such configurations cannot be stable, 
the entanglement entropy of $B_+$ has no contribution from the island. 

The position of $a_+$ can be determined by the same procedure 
to the case of $R$ after the Page time, since interaction 
between the right and left wedges of the bifurcate horizon 
can be ignored for $R$ with sufficiently large $t_b$, 
while all the twist operators are located in the right wedge for $B_+$. 
Thus, the quantum extremal surface is located almost at the same position 
to the case of $R$ after the Page time, 
which is given by \eqref{long-a} for $b \gg r_h$ and \eqref{short-a} for $b \gtrsim r_h$, respectively. 
The entanglement entropy of $B_+$ is approximately 
a half of that of $R$ (or equivalently $B$) after the Page time,%
\footnote{%
To be more precise, the quantum extremal surfaces for $B$ are located slightly 
inside than that for $B_\pm$ because of the interaction between the left and right wedges, 
and hence, the entanglement entropy of $B_\pm$ is slightly larger than a half of that of $B$. 
} 
\begin{align}
 S 
 &= 
 \frac{\pi r_h^2}{G_N} + \frac{\pi b^2}{G_N}
 + \frac{c}{12} 
 \left[
 \log \left(\frac{16 r_h^3 (b-r_h)^2}{b} \right) + \frac{b-r_h}{r_h}  
 \right] \ , 
 \label{long-SB}
\end{align}
for $b\gg r_h$, and 
\begin{align}
 S 
 = 
 \frac{\pi r_h^2}{G_N} + \frac{\pi b^2}{G_N} 
 - \pi  \kappa \, c \, \frac{r_h}{b-r_h} \ , 
\end{align}
for $b \gtrsim r_h$. 
Here, we do not consider the evaporation of the black hole due to the Hawking radiation 
since the vacuum state is the Hartle-Hawking vacuum, 
and then, the entanglement entropy of $B$ after the Page time gives its upper bound. 
Thus, the entanglement entropy of $B_\pm$ satisfies 
the subadditivity condition with that of $B = B_+\cup B_-$.


\subsection{Entanglement entropy of multiple regions}


\subsubsection*{Entanglement entropy of $R_\pm\cup B_\pm$}

\begin{figure}
\begin{center}
\includegraphics[scale=0.3]{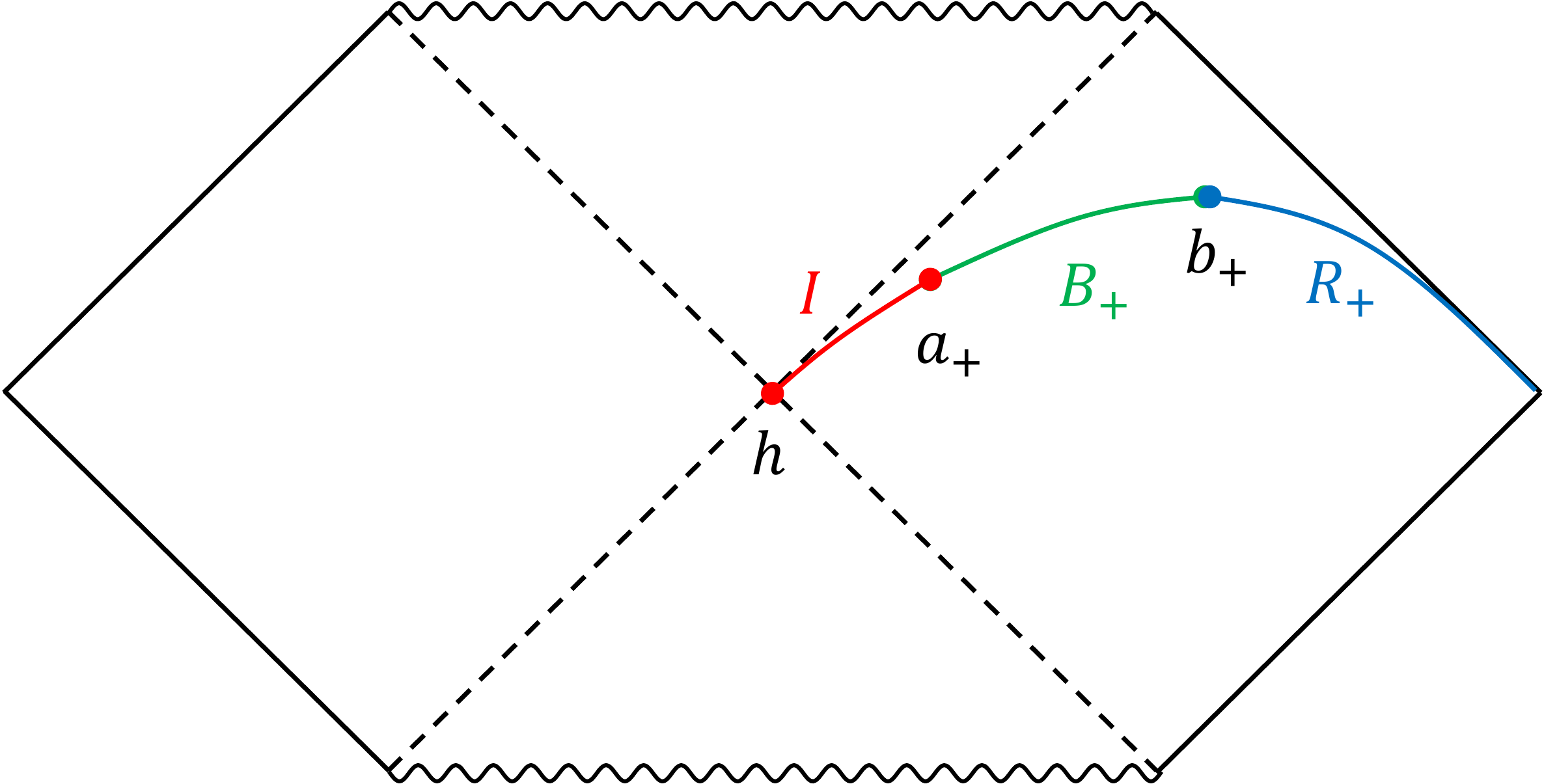}
\hspace{24pt}
\includegraphics[scale=0.3]{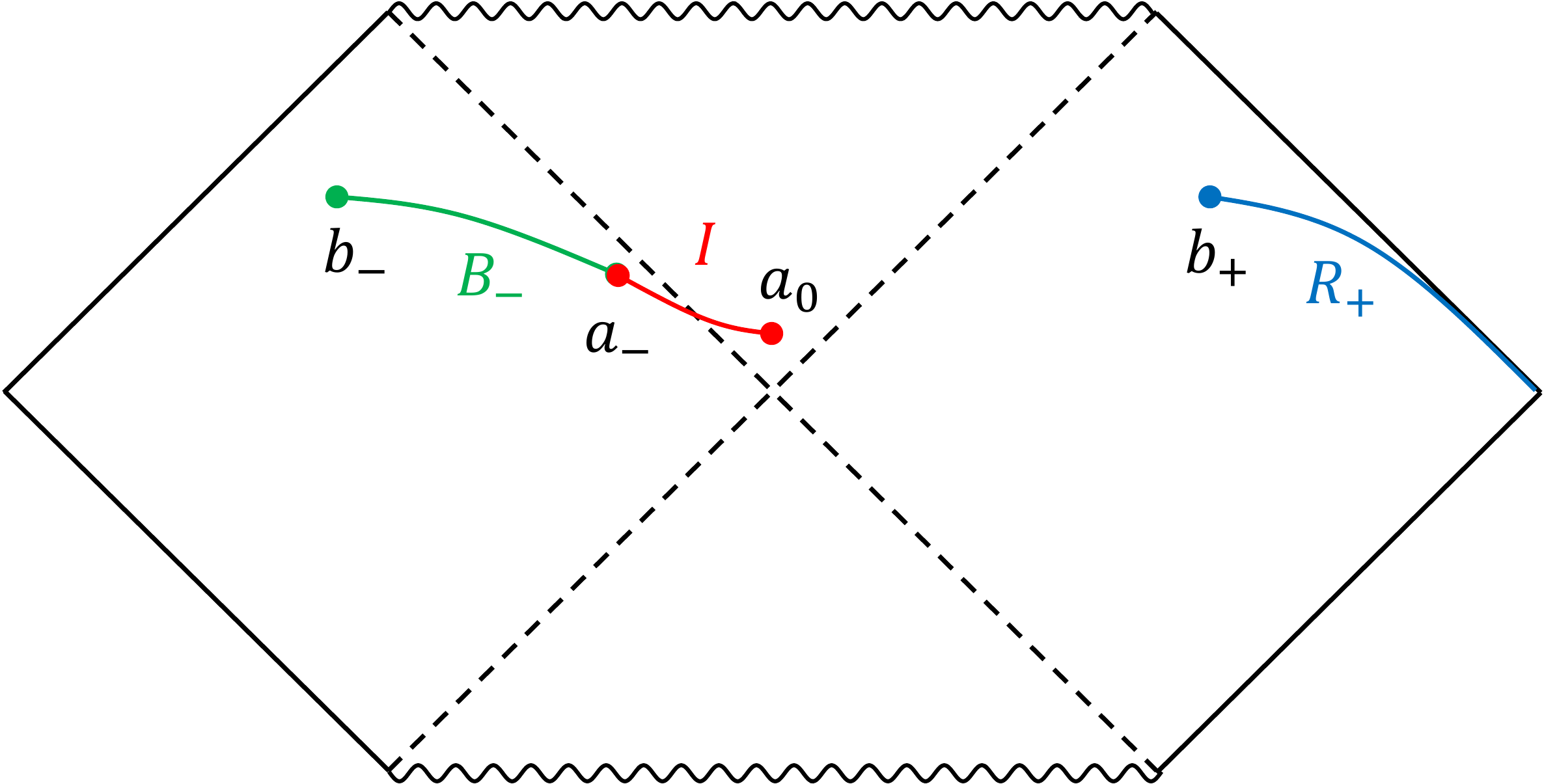}
\caption{%
The effective region of $R_+\cup B_+$ (left) and 
that of $R_+\cup B_-$ (right) include a half of the island. 
For $R_+\cup B_+$, the quantum extremal surface is located at 
the bifurcation sphere of the event horizon, $h$. 
The region is simply a time slice in the right wedge, but it includes the island $I$. 
Thus the effective region of $R_+\cup B_+$ is given by $R_+\cup B_+\cup I$, 
but the island $I$ is a half of that for $R$ or $B$. 
The island for $R_+ \cup B_-$ is also a half of that of $R$ or $B$, 
but includes a region inside the event horizon. 
}\label{fig:BH-half}
\end{center}
\end{figure}

Next, we consider the entanglement entropy of unions of two regions. 
We first consider the union of $R_+$ and $B_+$ (Fig.~\ref{fig:BH-half}). 
It is expected that the union of the Hawking radiation and the black hole 
in the right wedge covers the entire of a time-slice in the right wedge. 
The region is defined as the outside of the inner boundary 
which is given by the quantum extremal surface. 
The quantum extremal surface in this case is, in fact, 
located at the bifurcation sphere of the event horizon, 
$U=V=0$. 
This can be understood as follows. 

In this case, the region $R_+$ is continuously connected to $B_+$, 
and hence, no twist operators are inserted except for that at the quantum extremal surface. 
Thus, the quantum extremal surface has no quantum correction from the matter part, 
and hence, it is simply given by the extremum of the area of the surface. 
The extremal surface should be a local minimum with respect to variations in spatial directions, 
while it is a local maximum with respect to timelike variations. 
The bifurcation sphere of the event horizon, $U=V=0$, in fact, satisfies this condition. 

We can see that the quantum extremal surface is located at $U=V=0$,
also by taking the $b\to\infty$ limit of $B_+$. 
In this limit, $B_+$ covers almost all region outside the quantum extremal surface, 
and can be interpreted as $B_+\cup R_+$. 
Then, the position of the quantum extremal surface 
\eqref{short-a} approaches to the event horizon, $a\to r_h$, in $b\to\infty$. 
It is nothing but $U=V=0$ since the Schwarzschild time at the extremal surface $t_a$ is finite. 

Thus, the quantum extremal surface for $R_+\cup B_+$ is 
nothing but the bifurcation sphere of the event horizon. 
By comparing with the quantum extremal surface for $B_+$, 
the entanglement entropy contains contributions from the island, 
which is the region between the quantum extremal surface for $B_+$ 
and the bifurcation sphere of the event horizon 
($a_+$ and $h$ in Fig.~\ref{fig:BH-half}(left), respectively). 
The island for $R_+\cup B_+$ is a half of that for $R = R_+ \cup R_-$ after the Page time, 
or that for $B = B_+ \cup B_-$ before the Page time. 
In this case, the island does not include the interior of the event horizon. 

By introducing the IR cut-off of the spacetime for $R_+$, 
the entanglement entropy is calculated as 
\begin{equation}
 S = \frac{\pi r_h^2}{G_N} + \frac{c}{6} \log \Lambda \ . 
\end{equation}
This entanglement entropy obviously satisfies 
the subadditivity condition with those of two subregions 
because of the absence of the area term at $b_+$. 

Note that $b\to\infty$ limit of \eqref{long-SB} gives a slightly different result 
since it is calculated by using a different IR-cut-off scheme. 
The entanglement entropy of $B_+$ always contains the contribution from the outer boundary at $r=b$. 
However, the region $R_+$ is defined by introducing the IR-cut off of the spacetime in this paper, 
and has no contribution from the outer boundary. 
Thus, $b\to\infty$ limit of $B_+$ does not reproduce $R_+\cup B_+$ 
but has a slightly different IR cut-off terms. 
In order to check the subadditivity condition, 
we should choose the consistent IR cut-off with $R_+$. 

\subsubsection*{Entanglement entropy of $R_\pm\cup B_\mp$}

Next, we consider the union of $R_+$ and $B_-$ (or $R_-$ and $B_+$), 
which is the union of the Hawking radiation in one-side and the black hole in the other side. 
Definitions of the region $R_+$ and $B_-$ are the same to that for a single region, 
but the position of the quantum extremal surface, which gives the inner boundary of $B_-$, 
is different from the case of the single region of $B_-$. 

For $R_+\cup B_-$, two twist operators are inserted at $r = b$ 
in the both sides of the event horizon, and hence, 
the configuration (positions of the twist operators) is symmetric under the exchange of $U$ and $V$. 
Thus, the quantum extremal surface is on $U=V$ surface, and for $t_b>0$, 
is placed in the future wedge inside the future horizon, 
namely $U>0$ and $V>0$. 
Thus, the entanglement entropy has contributions from the island, 
which is a half of that for $R$ or $B$. 
In contrast to the island for $R_+\cup B_+$, 
the island for $R_+\cup B_-$ includes the region inside the event horizon. 

The position of the quantum extremal surface and 
the entanglement entropy can be calculated 
in a similar fashion to those for the Hawking radiation. 
For $b\gg r_h$, or the distance among the quantum extremal surface and 
the other boundaries of the region at $r = b$ is sufficiently large, 
the extremal surface is placed at 
\begin{equation}
 a = 
\begin{cases}
 \displaystyle
 r_h - \frac{c^2 G_N^2}{36 \pi^2 (b-r_h)r_h^2} 
 e^{ \frac{r_h-b}{r_h}} \sinh^2\left(\frac{t_b}{2 r_h}\right) \ , 
 & 
 \displaystyle
 t_b \ll r_h \log \frac{r_h^2}{G_N} \ , 
 \\
 \displaystyle
 r_h - \frac{c G_N}{12\pi r_h} \ , 
 &
 \displaystyle
 t_b \gg r_h \log \frac{r_h^2}{G_N} \ , 
\end{cases}
\end{equation}
and the entanglement entropy is calculated as  
\begin{equation}
 S = \frac{\pi r_h^2}{G_N} + \frac{2 \pi b^2}{G_N} + \frac{c}{6} \log \Lambda
 + 
\begin{cases}
 \displaystyle
 \frac{c}{6} \log \left(\frac{2 r_h^3(b-r_h)}{b \cosh^2 \left(\frac{t_b}{2 r_h}\right)}\right) \ , 
 & 
 \displaystyle
 t_b \ll r_h \log \frac{r_h^2}{G_N} \ , 
 \\
 \displaystyle
 \frac{c}{6}\left[-\frac{b}{r_h} + \log\left(\frac{2 c G_N r_h^2}{3 b \pi}\right)\right] \ , 
 &
 \displaystyle
 t_b \gg r_h \log \frac{r_h^2}{G_N} \ . 
\end{cases} 
\end{equation}

If the quantum extremal surface and the other boundaries of the region at $r = b$ 
are sufficiently close to each other, the extremal surface is placed at 
\begin{equation}
 a = 
\begin{cases}
 \displaystyle
 r_h - \frac{c^2 \kappa^2 G_N^2}{(b-r_h)^3} \sinh^2\left(\frac{t_b}{2 r_h}\right) \ , 
 & 
 \displaystyle
 t_b \ll r_h \log \frac{r_h^2}{G_N} \ , 
 \\
 \displaystyle
 r_h - \left(\frac{c^2 \kappa^2 G_N^2}{4 (b-r_h)} \right)^{1/3} e^{- \frac{t_b}{3 r_h}} \ , 
 &
 \displaystyle
 t_b \gg r_h \log \frac{r_h^2}{G_N} \ , 
\end{cases}
\label{S(R+B-)-long}
\end{equation}
and the entanglement entropy becomes 
\begin{equation}
 S = \frac{\pi r_h^2}{G_N} + \frac{2 \pi b^2}{G_N} + \frac{c}{6} \log \Lambda
 + 
\begin{cases}
 \displaystyle
 -\frac{2 \pi c \kappa r_h}{b-r_h} \ , 
 & 
 \displaystyle
 t_b \ll r_h \log \frac{r_h^2}{G_N}  \ , 
 \\
 \displaystyle
 -\left(\frac{2 \pi^3 c^2 \kappa^2 r_h^3}{G_N (b-r_h)} \right)^{1/3} e^{-\frac{t_b}{3 r_h}} 
 - \frac{2 \pi c \kappa r_h}{b-r_h} \ , 
 &
 \displaystyle
 t_b \gg r_h \log \frac{r_h^2}{G_N} \ . 
\end{cases} 
\label{S(R+B-)-short}
\end{equation}

Thus, the entanglement entropy satisfies the subadditivity condition with those of two subregions. 
It is always smaller than the sum of those of two subregions for $b\gg r_h$, 
while the expression \eqref{S(R+B-)-short} approaches to the same value to the sum of those of two subregions. 
It should be noted that the expression \eqref{S(R+B-)-short} is valid only if the distance 
among the quantum extremal surface $a_0$ and the other boundaries of the region $b_\pm$ 
are sufficiently close to each other. 
However, the distance becomes larger and larger as $t_b$ increases, 
since the quantum extremal surface is located in the future wedge inside the horizon. 
For sufficiently large $t_b$, the entanglement entropy 
is given by \eqref{S(R+B-)-long} even for $b\gtrsim r_h$. 

\subsubsection*{Entanglement entropy of unions of three regions}

Unions of three regions are complements of the other single region. 
Thus, the entanglement entropy is also the same to that of the single region, 
\begin{align}
 S(R_+\cup B_+ \cup B_-) &= S(R_-\cup B_+ \cup B_-) = S(R_+) = S(R_-) \ , 
\\
 S(B_+\cup R_+ \cup R_-) &= S(B_-\cup R_+ \cup R_-) = S(B_+) = S(B_-) \ . 
\end{align}
It is also straightforward to see that 
the entanglement entropy satisfies the strong subadditivity condition; 
\begin{equation}
 S(A\cup B \cup C) + S(A) \leq  S(A\cup B) + S(A\cup C) \ , 
\end{equation}
by using the entanglement entropy which we have calculated so far.  
Thus, the island prescription gives consistent entanglement entropy, 
although the position of the island depends on the combinations of the regions.


\section{Islands and stretched horizon}\label{sec:stretch}


In the previous section, we argued that the region $I$ 
in the effective region of the black hole before the Page time 
can be interpreted as the island, in the sense that 
it is involved by gravitational effects in the replica geometries. 
After the Page time, the entanglement entropy of the Hawking radiation 
can be written as the entanglement entropy of the effective region $R\cup I$, 
whereas what we are really calculating is the entanglement entropy of $R$. 
In a similar fashion, before the Page time, the entanglement entropy of the black hole 
is given by that of the effective region $B\cup I$, 
but we obtain this result by calculating the entanglement entropy of $B$, 
and then, the island $I$ is involved by gravitational effects. 

The inner boundary of $B$ (or equivalently $B_\pm$) is determined by the quantum extremal surface, 
implying that an inner half of $B$ is also involved by gravitational effects.%
\footnote{%
In fact, the inner boundary of the effective region $R_\pm\cup B_\pm$ is given by the quantum extremal surface, 
but the effective region contains a half of the island $I$. 
Since $R_\pm\cup B_\pm$ is almost equivalent to $B_\pm$ in $b\to\infty$, 
the region $B_\pm$ includes a region which can be interpreted as the island in this sense. 
}
Thus, the region $B$ can be further separated into 
the essential region $B'$ and the ``hidden island'' $I'$ (Fig.~\ref{fig:BH-crit}(right)). 
The entanglement entropy of the black hole is identified with that of $B'$, 
but the region $B'$ is extended by gravitational effects and 
becomes effectively $B\cup I$ before the Page time and $B$ after the Page time. 

%

%
%
In the case of the AdS spacetime with an auxiliary system in the flat spacetime, 
the entanglement entropy of the black hole is identified with 
that in a region in the auxiliary flat space (Fig.~\ref{fig:AdSBH}). 
Thus the region in the auxiliary spacetime corresponds to $B'$. 
The region inside the AdS side appears as a consequence of gravitation, 
and hence corresponds to $I'$. 
In fact, the inner boundary of $I'$ is given by 
the quantum extremal surface $a_\pm$, as is the region in the AdS side in the case of the AdS-BH. 
In the case of AdS, the boundary between $B'$ and $I'$ is nothing but 
the interface between the AdS spacetime and the auxiliary spacetime. 


In the case of the asymptotically flat spacetime, 
the boundary between $B'$ and $I'$ can be found by taking the limit 
in which $B$ disappear, or equivalently, $R$ merges with $I$. 
The radius at the quantum extremal surface \eqref{short-a} increases 
as the radius at $b_\pm$ decreases, implying that 
they meet at some radius which is identified with the boundary between $B'$ and $I'$. 
In fact, in the case of AdS, both $a_\pm$ and $b_\pm$ approach to 
the interface between the AdS spacetime and the auxiliary flat spacetime. 
In this section, we will show that the quantum extremal surface becomes unstable for $b=b_c$, 
implying that the region $R$ and the island $I$ becomes continuous with each other. 
Thus, the region $B$ can be separated into $I'$ and $B'$, 
which are placed in $a\leq r < b_c$ and $b_c \leq r \leq b$, respectively.%
\footnote{%
Note that the surface at $r=b_c$ is included in $B'$ 
since it corresponds to the interface in the case of AdS with auxiliary spacetime 
and would have some information of the black hole.
}
Thus, the information of the black hole and the Hawking radiation 
can be read off from $B'$ and $R$, respectively. 
In fact, $R\cup I$ covers entire spacetime for $b=b_c$, 
implying that $R$ has all the information of the total system in this case. 

\begin{figure}
\begin{center}
\includegraphics[scale=0.5]{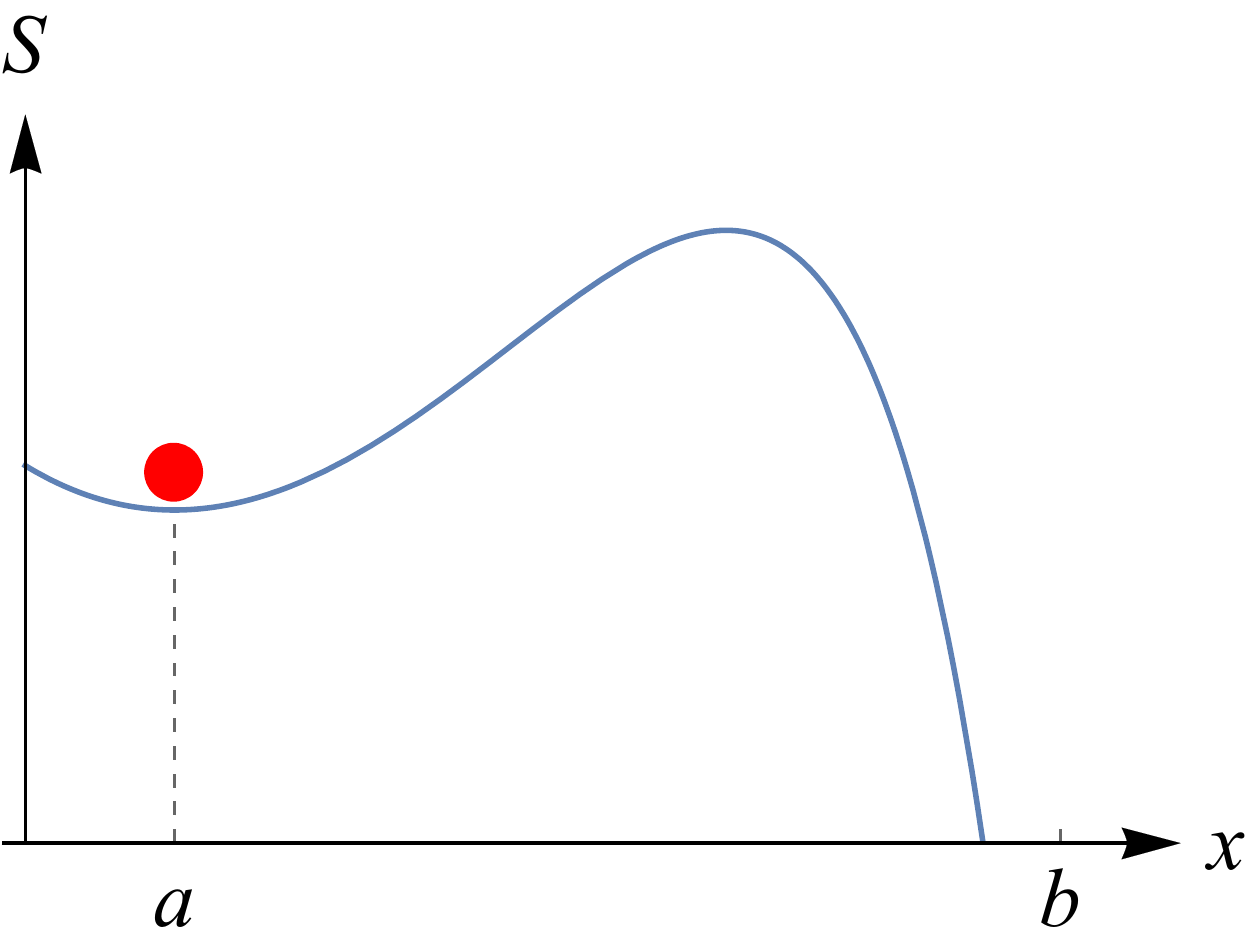}
\raisebox{60pt}{
\begin{tabular}{c}
$b\to b_c$ \vspace{4pt}\\
{\Huge$\longrightarrow$}
\end{tabular}
}
\includegraphics[scale=0.5]{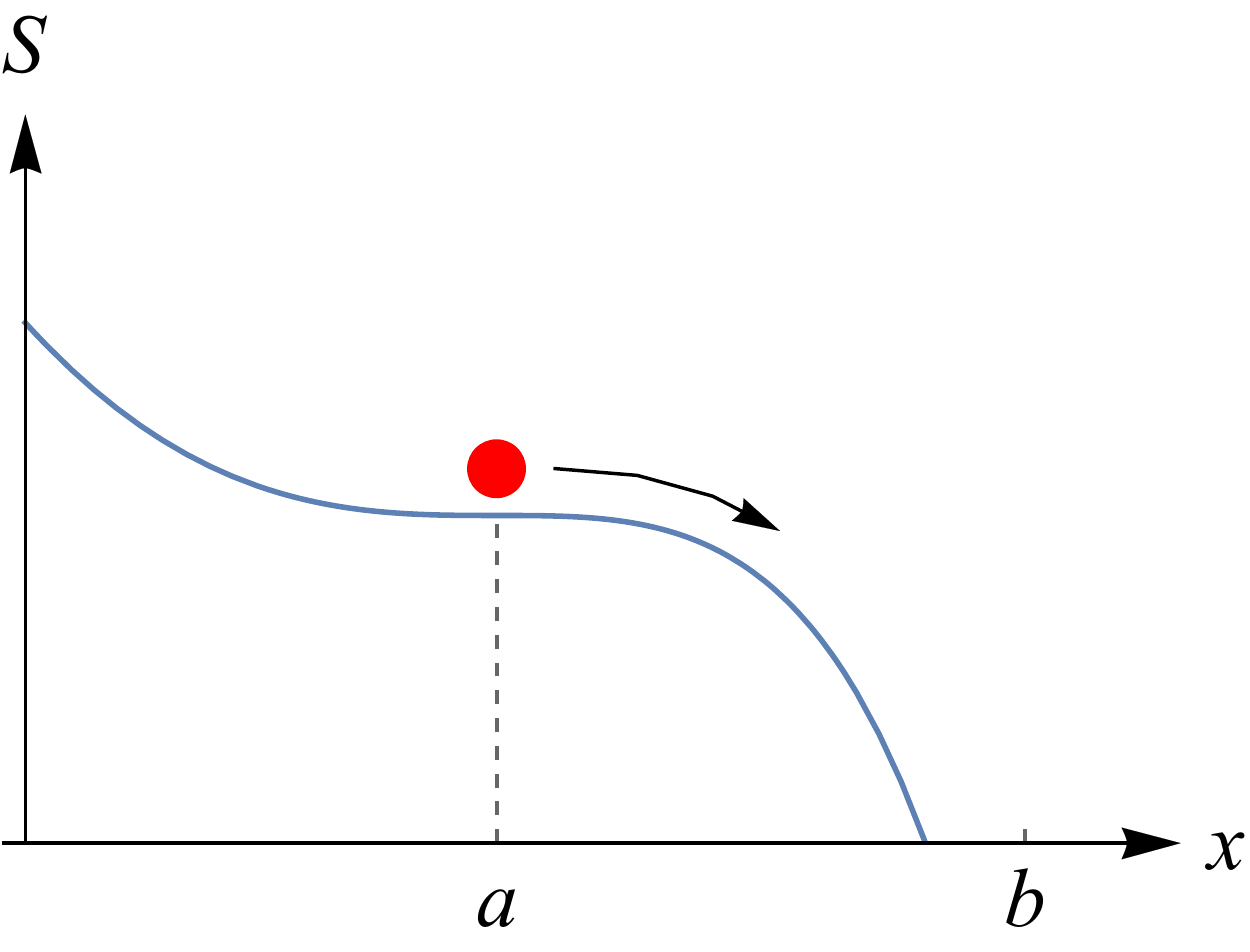}
\caption{%
Phase transition of the quantum extremal surface 
when the boundary $b_\pm$ between $R$ and $B$ approaches to the quantum extremal surface, $a_\pm$. 
The graphs show the entanglement entropy of $R$ as a function of the position 
of the quantum extremal surface $x$, where $x = \sqrt{\frac{r-r_h}{r_h}}$\,. 
If the boundary $b_\pm$ is not very close to $a_\pm$, 
the entropy has a local minimum and local maximum (left). 
When, the boundaries $b_\pm$ have sufficiently approached to $a_\pm$, namely at $b=b_c$, 
the local minimum merges to the local maximum, and then, 
the entropy has no local minimum (right). 
}\label{fig:critical}
\end{center}
\end{figure}

Now, we calculate $b_c$, which is the radius of $b_\pm$ when $a=b$.  
The radius of the quantum extremal surface is given by 
the saddle point of the entanglement entropy 
\begin{equation}
 \partial_a S = 0 \ , 
\end{equation}
which can be expressed by in terms of 
$x = \sqrt{ \frac{a-r_h}{r_h}}$ and $y = \sqrt{ \frac{b-r_h}{r_h}}$ as 
\begin{align}
 x \left(y-x\right)^3 &= \frac{ \kappa c \, G_N}{2r_h^2} \ .
\end{align}
This equation has two solutions for $0 < x < y$, or equivalently, $r_h < x < b$. 
The solution with larger $x$ is a local maximum of the entanglement entropy. 
That with smaller $x$ is a local minimum of the entanglement entropy, 
which is a saddle point of the path integral (Fig.~\ref{fig:critical}(left)). 
The local minimum and local maximum approaches to each other 
as $y$ becomes smaller, or equivalently, as $b$ approaches to $r_h$. 
They merge when the radius of the quantum extremal surface 
and that of the boundary between $R$ and $B$ are given by 
\begin{align}
 a &= a_c = r_h + \sqrt{ \frac{27\kappa c\,G_N}{2}} \ , 
&
 b &= b_c = r_h + 24\sqrt{6\kappa c\,G_N} \ .  
\label{bc}
\end{align}
Note that $a=a_c$ is no longer the local minimum 
though $\partial_a S = 0$ there (Fig.~\ref{fig:critical}(right)). 
This result implies that the quantum extremal surface becomes unstable 
before $a_\pm$ and $b_\pm$ coincide with each other. 

The instability implies that the quantum extremal surface decays into $a=b$ for $b\leq b_c$.%
\footnote{%
Naively, no island appears for $b \leq b_c$ 
since it is no longer a local minimum of the entanglement entropy. 
However, 
the region $R$ with $b\leq b_c$ does not contain the information of the island $I$ 
even after the Page time, if there is no island. 
This is inconsistent with the fact that the the region $R$ with $b>b_c$ 
has the information of the island, 
since $R$ with $b>b_c$ is included in $R$ with $b\leq b_c$. 
In fact, the instability of the quantum extremal surface 
implies only that no island with $a\neq b$. 
} 
The entanglement entropy plays a similar role to the potential energy or free energy.  
The quantum extremal surface at $a\neq b$ is a ``meta-stable vacuum'' for $b>b_c$, 
but there is no ``barrier'' to decay into the ``true vacuum'' for $b\leq b_c$. 
The ``true vacuum'' is not well-defined within the framework of the semi-classical gravity, 
and hence, it is unclear whether there is such a saddle point in the path integral. 
However, the instability of the quantum extremal surface implies that 
there is another vacuum in this direction at least for $b\leq b_c$. 

Thus, the instability indicates that the island, 
which is the branch cut in the replica geometries, will be extended toward $b_\pm$. 
The negative divergence of the entanglement entropy at $a=b$ 
appears because two twist operators have approached below the UV cut-off scale, $\ell_p = G_N^{1/2}$. 
It implies that we should not take $a=b$ exactly because of the UV-cut-off. 
The entanglement entropy approaches zero as $a$ approaches to $b$, 
and becomes zero at some point where the distance between $a$ and $b$ 
is of the same order to the UV cut-off, $\ell_p$. 
Then, it should be interpreted that $R\cup I$ is a single continuous region, and $B$ disappears. 
Thus, we expect that the entanglement entropy becomes zero when $b = b_c$, 
although it cannot be seen directly because physics at the UV cut-off scale 
cannot be described by the semi-classical gravity. 


Although $b_c - r_h = \mathcal O(\ell_p)$, the proper distance between 
the event horizon and the critical radius at $r = b_c$ is 
of $\mathcal O(\sqrt{r_h \ell_p})$, which is much longer than the UV cut-off, $\ell_p$. 
Although the instability implies that the entanglement entropy is zero for $b=b_c$, 
the entanglement entropy of the region $R$ after the Page time still has 
\begin{equation}
 S = \frac{2 \pi r_h^2}{G_N} + \mathcal O(G_N^{-1/2}) \ , 
\end{equation}
if $b$ is even slightly larger than $b_c$. 
Thus, the entanglement entropy as a function of $b$ is discontinuous and jumps to zero at $b=b_c$. 
In the same way, the entanglement entropy of the black hole $B_+$ (or $B_-$) 
approximately equals to the Bekenstein-Hawking entropy as long as $b>b_c$, 
but the region $B_+$ itself vanishes when $b=b_c$.
This implies that the information of the black hole is localized 
on the surface at $r = b_c$, or more precisely, 
within the UV cut-off scale around the radius $r = b_c$. 
It should be noticed that the argument above is valid for $B_+$ 
whichever before or after the Page time. 
Therefore, the islands in the Schwarzschild spacetime 
indicates that the information of the black hole is 
localized on the stretched horizon at $r = b_c$. 
The information is not cloned from the collapsed matters 
but would be reproduced by a similar mechanism to decode 
the information in the island $I$ from the Hawking radiation.


\section{Discussions}\label{sec:discussion}

In this paper, we found an instability of the quantum extremal surface for $b = b_c$, 
where the boundary between the regions $R$ and $B$ are sufficiently close to the horizon. 
Assuming that this instability implies that the quantum extremal surface 
decays into $a = b$, or equivalently, the island $I$ and the region $R$ merge with each other, 
the information of the black hole should be localized on the stretched horizon at $r = b_c$. 
Hereafter, we call the region $R$ and the island $I$ with $b=b_c$ as $R_c$ and $I_c$, 
and use the term $R$ and $I$ only those for $b>b_c$. 
The entanglement entropy of the region $R_c$ is associated with the island $I_c$ 
which is the complement of the region $R_c$ itself, and hence, 
the entanglement entropy is zero. 
Thus, the region $R_c$, or equivalently the region $r\geq b_c$,  
contains the all the information of the total system. 
When we divide the system into the Hawking radiation and the black hole, 
the region $R_c$ is further divided by boundaries $b_\pm$ at $b > b_c$ 
into the regions $R$ with $b>b_c$ and $B'$, namely, $R_c = R\cup B'$.%
\footnote{%
We also have a similar relation for the island $I_c = I\cup I' = I\cup I'_+ \cup I'_-$. 
The other relations between the regions are summarized as follows: 
$B = B' \cup I'$, $B'= R_c\cap B$ and $I' = B\cap I_c$. 
} 
Then, the information of the Hawking radiation is located in the region $R$, 
and the information of the black hole is located in the region $B'$. 
When we calculate the entanglement entropy of $B'$ after the Page time by using the replica trick, 
the hidden island $I'$ appears due to gravitational effects, 
and the region effectively becomes $B$. 
For the entanglement entropy of $B'$ before the Page time, 
the island $I$ also appears and the effective region becomes $B\cup I$. 
It depends on $b$ how to distinguish the Hawking radiation and black hole, 
but the size of $B'$ can be arbitrarily small. 
Since the entanglement entropy of $B'$ reproduces the Bekenstein-Hawking entropy as long as $b>b_c$, 
the information of the black hole is localized on the surface at $r=b_c$, 
which is interpreted as the stretched horizon. 

This structure is similar to the case of 
the AdS-BH spacetimes with auxiliary flat spacetime \cite{Almheiri:2019qdq}. 
In that case, we calculate the entanglement entropy of regions in the auxiliary flat spacetime, 
while regions in the AdS spacetime, or equivalently, 
the branch cuts in the AdS side 
is involved by the gravitation in replica geometries. 
By using the AdS/CFT correspondence, 
the information of the black hole can be encoded into the AdS boundary, or equivalently, 
the interface with the auxiliary flat spacetime. 
Since the AdS spacetime can be obtained by taking the near horizon limit of the black brane solution, 
the boundary of the near horizon region, or equivalently, the AdS boundary 
can be interpreted as the stretched horizon. 
Therefore, the island rule in the asymptotically flat spacetime has 
similar structures to that in the AdS spacetime; 
(i) only the stretched horizon and its exterior should be taken into account 
in the calculation of the entanglement entropy, 
and, (ii) the information of the black hole is localized at the stretched horizon. 
In fact, the correspondence between the black hole and 
the membrane (or fluid) on the stretched horizon is already known as 
the membrane paradigm \cite{Damour:1978cg,Damour:1982,Price:1986yy,Parikh:1997ma} 
from long before the AdS/CFT correspondence. 

It should be noted that we studied the island rule only 
at the leading order of the semi-classical approximation. 
Our result reproduces the information of the black hole only up to the higher order corrections. 
It is known in the framework of the membrane paradigm 
that the stretched horizon at a proper distance $L$ from the event horizon 
reproduces effects of the black hole up to fractional errors of $\mathcal O(L^2)$ \cite{Price:1986yy}. 
Since our stretched horizon is located at $L \sim \sqrt{r_h \ell_p}$, 
it reproduce physics of the black hole up to corrections of $\mathcal O(\ell_p/r_h)$. 
This would be related to the higher order corrections of the semi-classical approximation. 
Our stretched horizon is distinguishable from real membranes as 
the distance from the horizon is larger than the Planck length. 
The island rule implies that the information of the black hole 
is reproduced from bulk field configurations at the stretched horizon, 
as the information of the island is reproduced from the Hawking radiation. 
Our stretched horizon is not exactly the same to 
that in black hole complementarity \cite{Susskind:1993if,Stephens:1993an}. 
The stretched horizon of black hole complementarity is placed at 
a Planck distance from the event horizon so that 
distant observers see an apparently real membrane on it \cite{Susskind:1993if}, 
whereas our stretched horizon behaves as a membrane only approximately.%
\footnote{%
The stretched horizon at $L \sim \sqrt{r_h \ell_p}$ is also equivalent to 
that at a Planck distance up to $\mathcal O(\ell_p/r_h)$ errors, 
but is distinguishable from the real membrane because of the errors. 
} 
Our result is limited to the leading order of the semi-classical approximation, 
while black hole complementarity is proposed to store all the information of black hole below the Planck scale. 
In order to reproduce all the information of the black hole, 
we need to consider the higher order corrections to the semi-classical approximation. 
The position of the stretched horizon might be corrected and become closer to the horizon 
after taking the higher order corrections, 
since the region $R$ might naively be larger to reproduce more information. 
This is out of the scope of this paper and left for future studies. 

It is also interesting that the entanglement entropy of the region $R_c$ --- 
the stretched horizon and the region outside it --- is zero. 
In this case, the quantum state in $R_c$ is pure state and contains all the information of the total system. 
This also implies that there would be a unitary map between the quantum states at different time. 
Thus, the quantum state of $R_c$ at arbitrary time $t_b$ 
can in principle be reconstructed from that at any given time. 
By taking the limit of $t_b\to-\infty$, the region $R_c$ approaches to the past null infinity, 
and the causal future of $R_c$ covers all the interior of the future horizon. 
Thus, $S=0$ is consistent with the fact that the region $R_c$ contains the information of the island $I_c$.%
\footnote{%
Here, we also assumed that the state on the past horizon is the trivial. 
}
The quantum state of $R_c$ in $t_b\to-\infty$ limit 
can be reconstructed from the quantum state of $R_c$ at a given time, 
and then, the quantum state in the island $I$ can be obtained 
from the time evolution of the state of $R_c$ in $t_b\to-\infty$. 
Although this argument cannot be applied directly for $R$ with $b>b_c$ as it is not a pure state, 
the information of the island might be reproduced from $R$ 
by using a similar but more complicated procedure for the subsystem. 

\begin{figure}
\begin{center}
\includegraphics[scale=0.3]{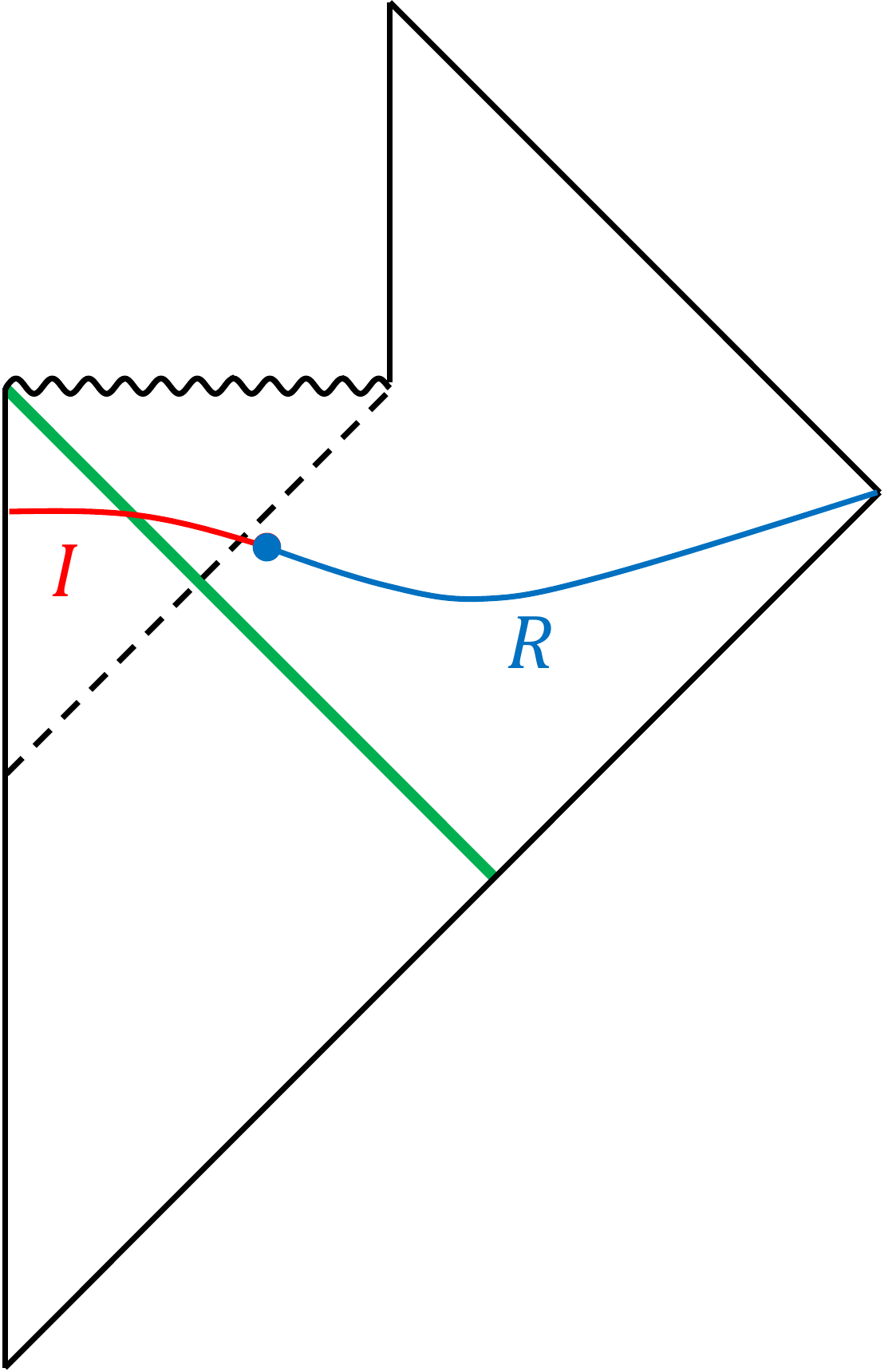}
\hspace{64pt}
\includegraphics[scale=0.3]{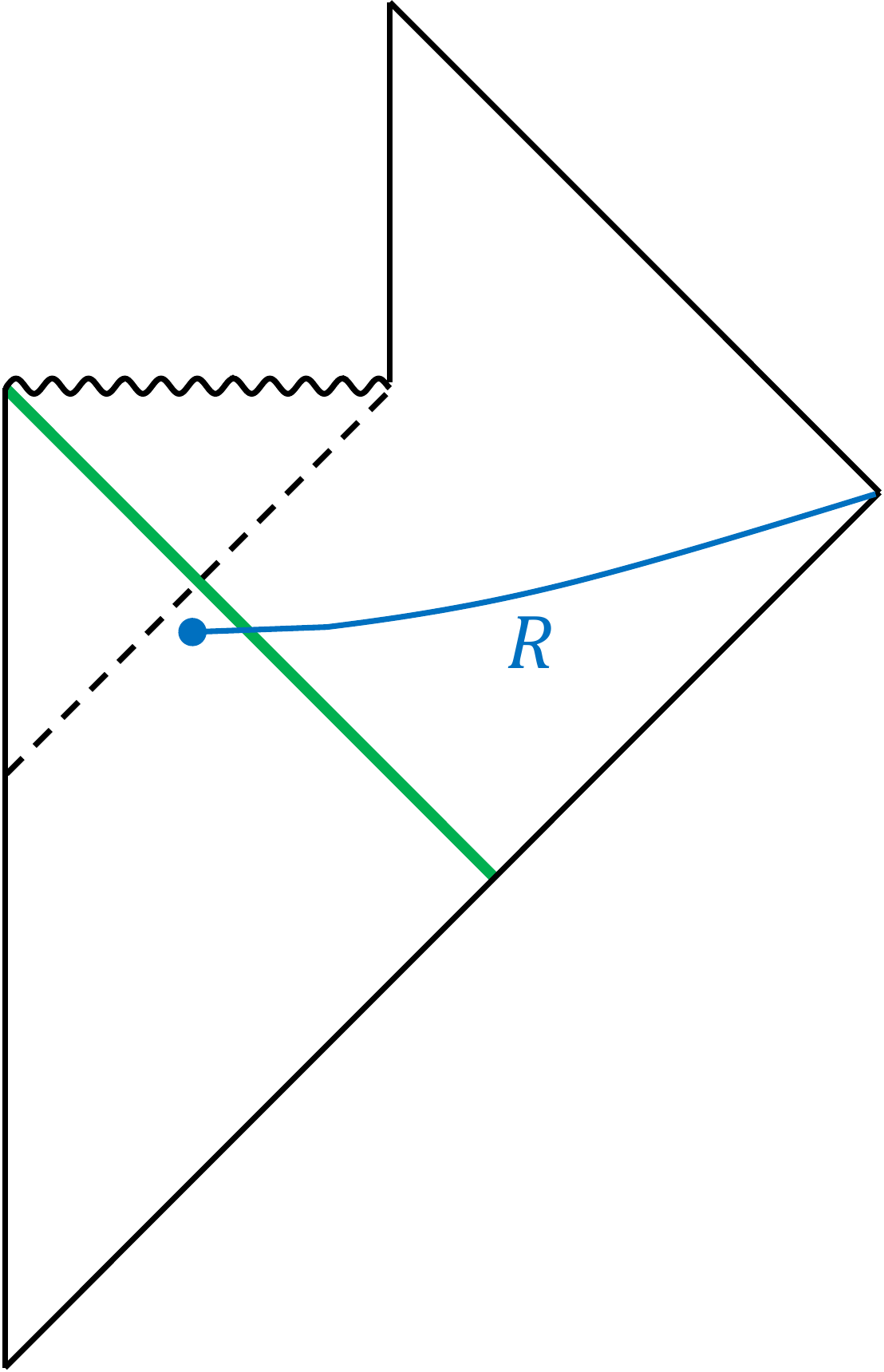}
\caption{%
The Penrose diagram of a black hole formed by a gravitational collapse of a thin null shell. 
The geometry outside the shell is approximately given 
by the Schwarzschild geometry, but the interior is the flat spacetime. 
The region $R$ for $b=b_c$ outside the shell is associated with 
the island $I$ which covers all the other region, and the entanglement entropy is zero (left). 
If the inner boundary of $R$ is inside the shell, the island would not appear 
since there is no unstable saddle point in the flat spacetime (right). 
However, the region $R$ still has the all the information of the initial state, 
provided that there is no other matter excitations other than the shell. 
}\label{fig:collapse}
\end{center}
\end{figure}

The argument on the island and the stretched horizon 
can be generalized into the case of a black hole which 
is formed by a gravitational collapse. 
For simplicity, we consider a thin null shell. 
The geometry outside the shell is given by the Schwarzschild geometry 
and the interior is the flat spacetime (Fig.~\ref{fig:collapse}). 
The position of the quantum extremal surface is determined 
so that the configuration with the island becomes a solution of 
the semi-classical Einstein equation for the replica geometries, 
and hence, depends mostly on the geometry near the inner boundary of the region $R$.%
\footnote{%
It should be noted that the position of the quantum extremal surface also depends on the vacuum state. 
The Unruh vacuum should be taken for the black hole formed by the gravitational collapse 
while the Hartle-Hawking vacuum is more appropriate for the eternal black hole. 
The island is mostly inside the horizon in the Unruh vacuum, 
but extends outside the horizon if $b$ is sufficiently close to the horizon 
(See Appendix~\ref{sec:position} for more details). 
This would be because the universal singular part of 
the correlation function of twist operators dominates in this case.
} 
After the gravitational collapse, or equivalently, 
when the inner boundary of the region $R$ is located outside the shell, 
the quantum extremal surface has the same unstable saddle point for some $b\,(=b_c)$
as in the case of the eternal black hole, implying the existence of 
the configuration in which the island merges with the region $R$. 
That configuration gives $S=0$ and hence would be most dominant even before the Page time. 
In a similar fashion to the case of the eternal black hole geometry, 
the information of the initial state can be read off from the region $R_c$, 
and the information of the island $I_c$ can also be reconstructed consistently. 

Before the gravitational collapse, when the inner boundary of the region $R$ is inside the shell, 
there is no unstable saddle point as the twist operator is on the flat spacetime, 
and hence, there is no island. 
In this case, the effective region does not cover the entire time-slice. 
However, the initial state can be reproduced from the information in $R$ 
since it covers the shell, provided that there is no other matter other than the shell. 
Thus, the initial state can always be reconstructed from the information in $R$. 

In this way, the island would imply that the unitarity is 
satisfied without the interior of the event horizon. 
This is a similar picture to other approaches to the information loss paradox, 
such as black hole complementarity \cite{Susskind:1993if,Stephens:1993an} 
or firewall \cite{Almheiri:2012rt}, but the information inside the event horizon 
is encoded into the stretched horizon and the Hawking radiation 
by a more complicated mechanism of quantum information in the gravitational system.

\section*{Acknowledgments}

The author would like to thank Koji~Hashimoto for helpful discussions. 
This work is supported in part by JSPS KAKENHI Grants No.~JP17H06462 and JP20K03930.


\begin{appendix}

\section*{Appendix}


\section{Analogy to holographic entanglement wedge}\label{sec:analogy}

In order to get a better understanding on islands, 
it is convenient to consider analogy to 
the entanglement wedge \cite{Czech:2012bh,Wall:2012uf,Headrick:2014cta} in holography, 
since the island is first proposed in the framework of holography. 
Here, by using this analogy, it is more appropriate to consider 
that the island appears even before the Page time and is included 
in the effective region of the black hole subsystem. 

In the evaporation of a black hole in the AdS spacetime, 
initially the entire bulk spacetime is the entanglemenet wedge of the boundary CFT. 
After (a part of) the Hawking radiation has escaped from the AdS spacetime, 
CFT corresponds only the other degrees of freedom than the Hawking radiation. 
The entanglement wedge of CFT is the region between the AdS boundary 
and the quantum extremal surface, and then, its complement is 
the entanglement wedge of the Hawking radiation, which is called islands. 
However, islands do not directly correspond to the Hawking radiation. 
In fact, there is a region of the Hawking radiation other than islands, 
if the junction with the auxiliary flat spacetime is considered. 

\begin{figure}[tb]
\begin{center}
\includegraphics[scale=0.25]{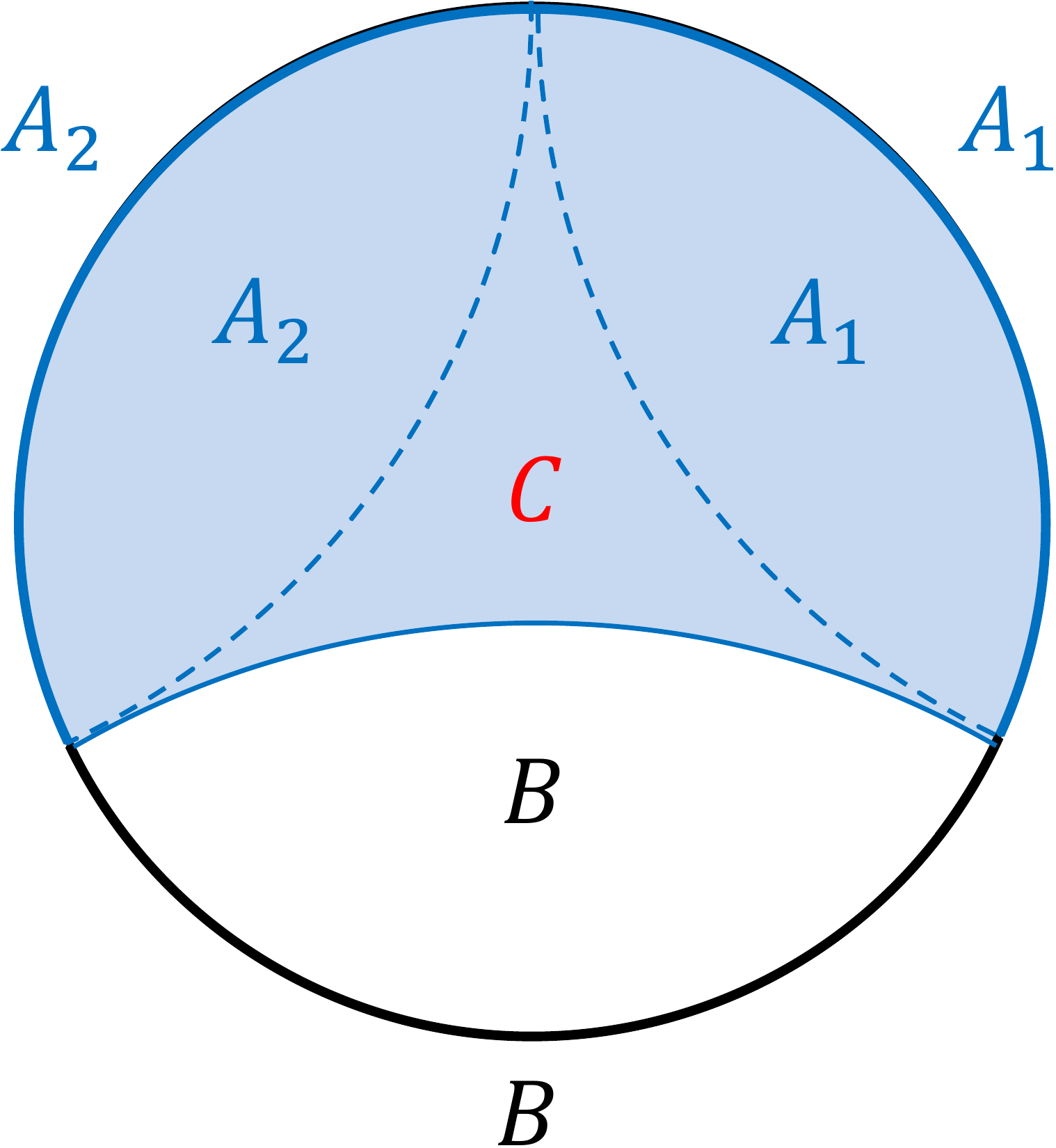}
\hspace{24pt}
\includegraphics[scale=0.25]{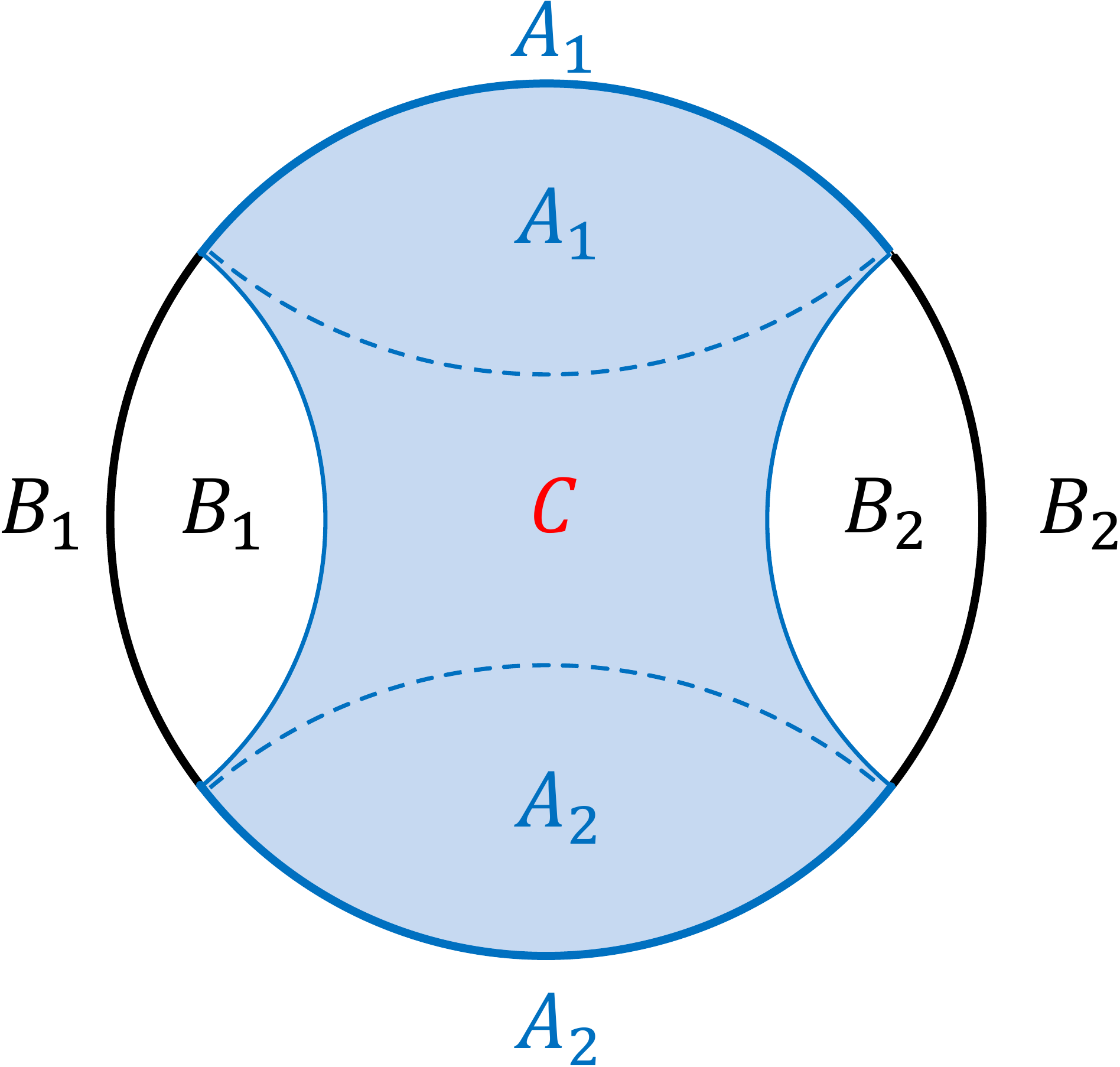}
\hspace{18pt}
\includegraphics[scale=0.25]{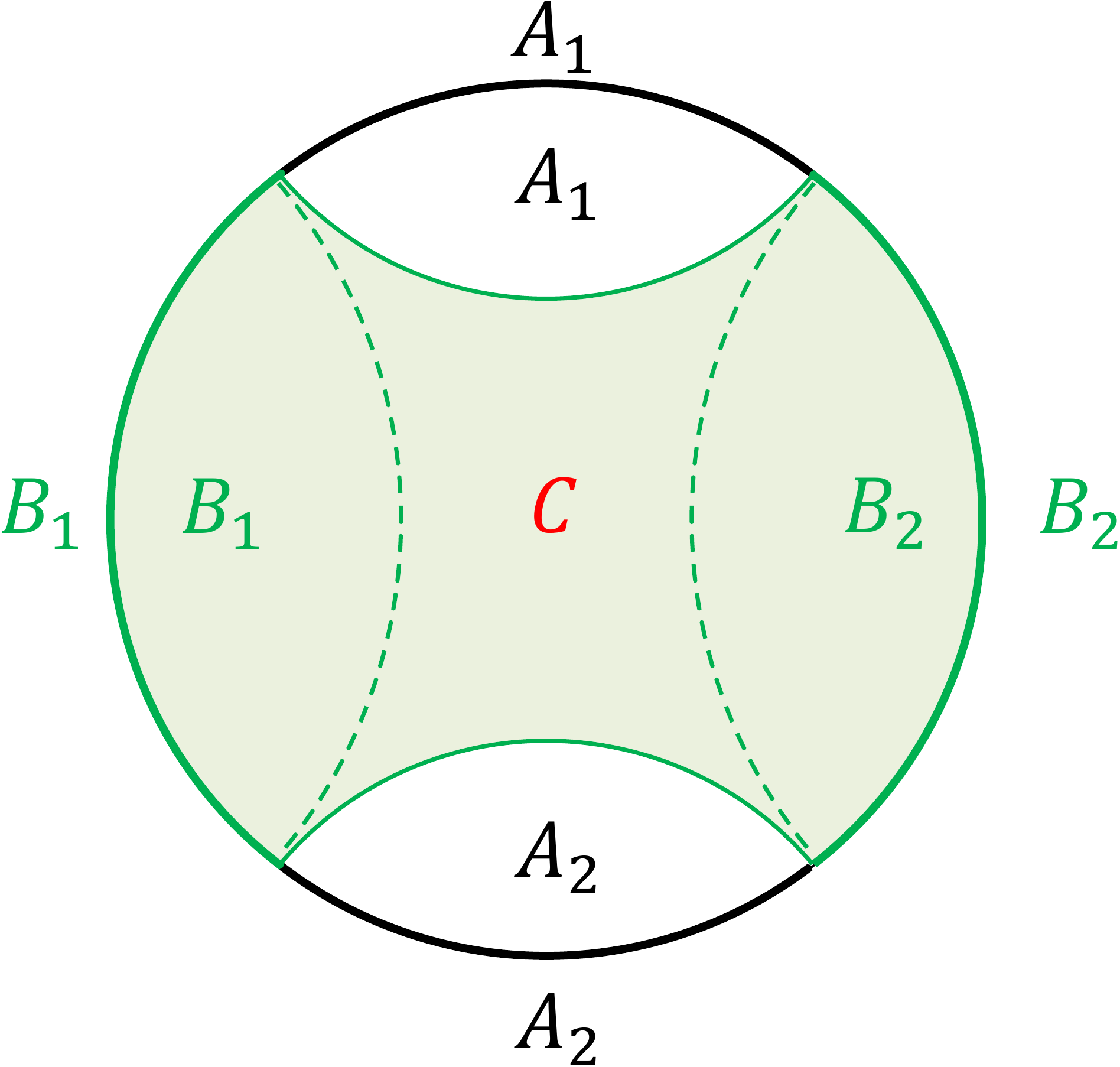}
\caption{%
If CFT is divided into three subsystems, $A_1$, $A_2$ and $B$, 
the bulk is divided into four regions: three entanglement wedges of $A_1$, $A_2$, $B$, 
and the other region $C$. The entanglement wedge of $B_1\cup B_2$ consists of $B_1$, $B_2$ and $C$ (left).  
If CFT is divided into four subsystems, $A_1$, $A_2$, $B_1$ and $B_2$,
the bulk is divided into four regions, $A_1$, $A_2$, $B_1$, $B_2$ and $C$. 
When the subsystem $A = A_1\cup A_2$ is larger than its complement $B = B_1\cup B_2$, 
the entanglement wedge of $A$ in the bulk includes $C$ (center). 
When the subsystem $B$ is larger, however, 
$C$ belongs to the entanglement wedge of $B$ and is not included in that of $A$ (right). 
This structure is similar to the regions of the Hawking radiation and black hole --- 
the regions $A_1$, $A_2$, $B_1$, $B_2$ and $C$ correspond to 
$R_+$, $R_-$, $B_+$, $B_-$ and $I$, respectively. 
}\label{fig:holo}
\end{center}
\end{figure}

In order to see what is the island, it is convenient to introduce 
auxiliary systems which correspond to the bulk Hawking radiation in the boundary side. 
The bulk counterpart of these auxiliary systems is the Hawking radiation itself 
which has already got out of the AdS spacetime. 
Here, for simplicity, we divide the Hawking radiation into two systems --- 
that in the right and left wedges, though 
it might be more appropriate to divide into a larger number of systems. 
Thus, we have divided the total system into three subsystems --- 
the Hawking radiation in the right wedge, that in the left wedge and the black hole.
\footnote{%
A simple model of the black hole evaporation is studied 
by introducing auxiliary subsystems of each Hawking quanta 
in \cite{Akers:2019nfi}. 
} 
Then, the bulk spacetime is divided into four regions --- 
the entanglement wedges of each subsystem and the other region (See Fig.~\ref{fig:holo}(left)). 
The region $C$ in Fig.~\ref{fig:holo} is not included in the entanglement wedge of $A_1$ or $A_2$, 
but is a part of the entanglement wedge of $A= A_1\cup A_2$. 
Thus, the region $C$ corresponds to the island. 

\begin{figure}[tb]
\begin{center}
\includegraphics[scale=0.25]{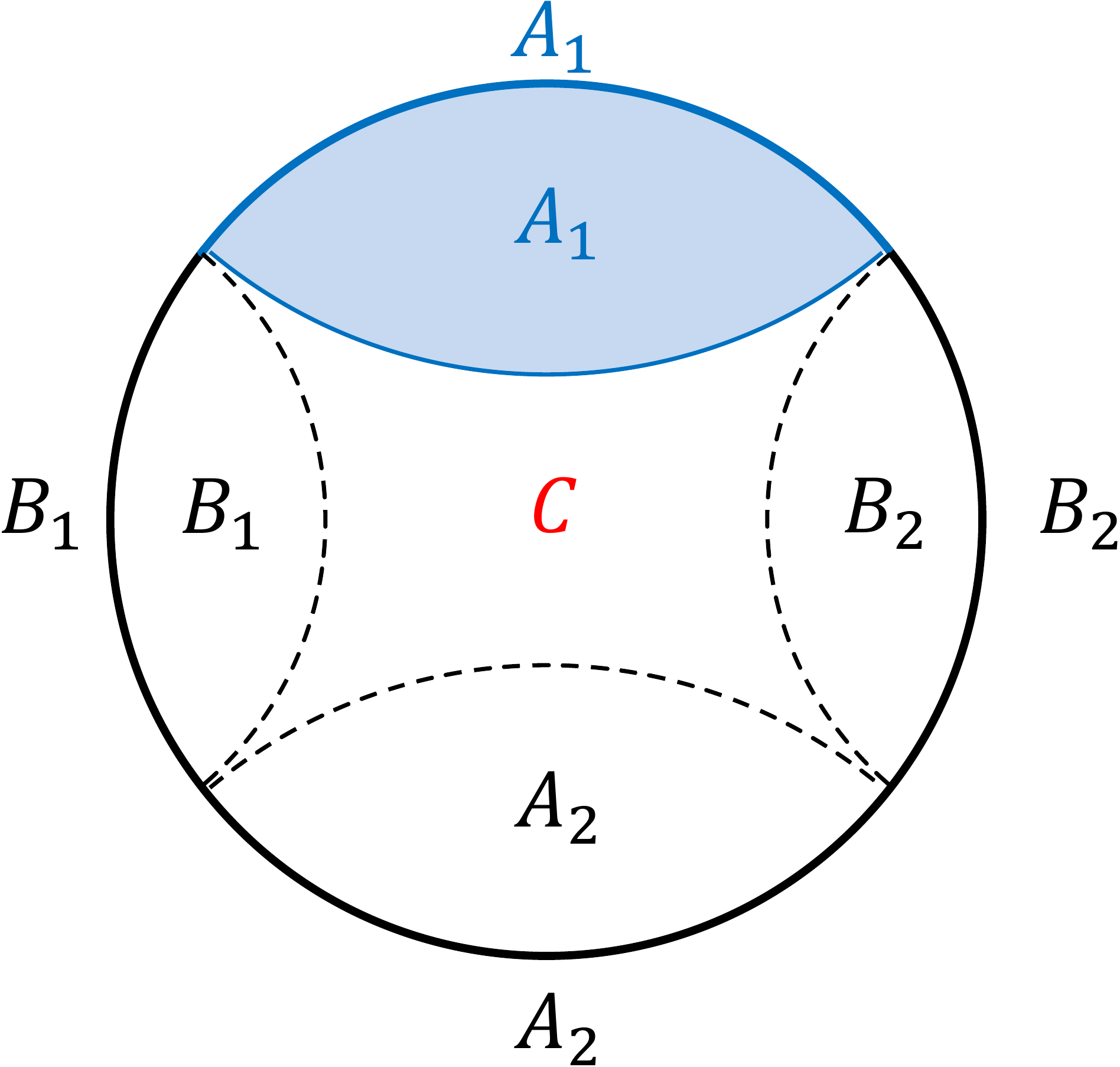}
\hspace{48pt}
\includegraphics[scale=0.25]{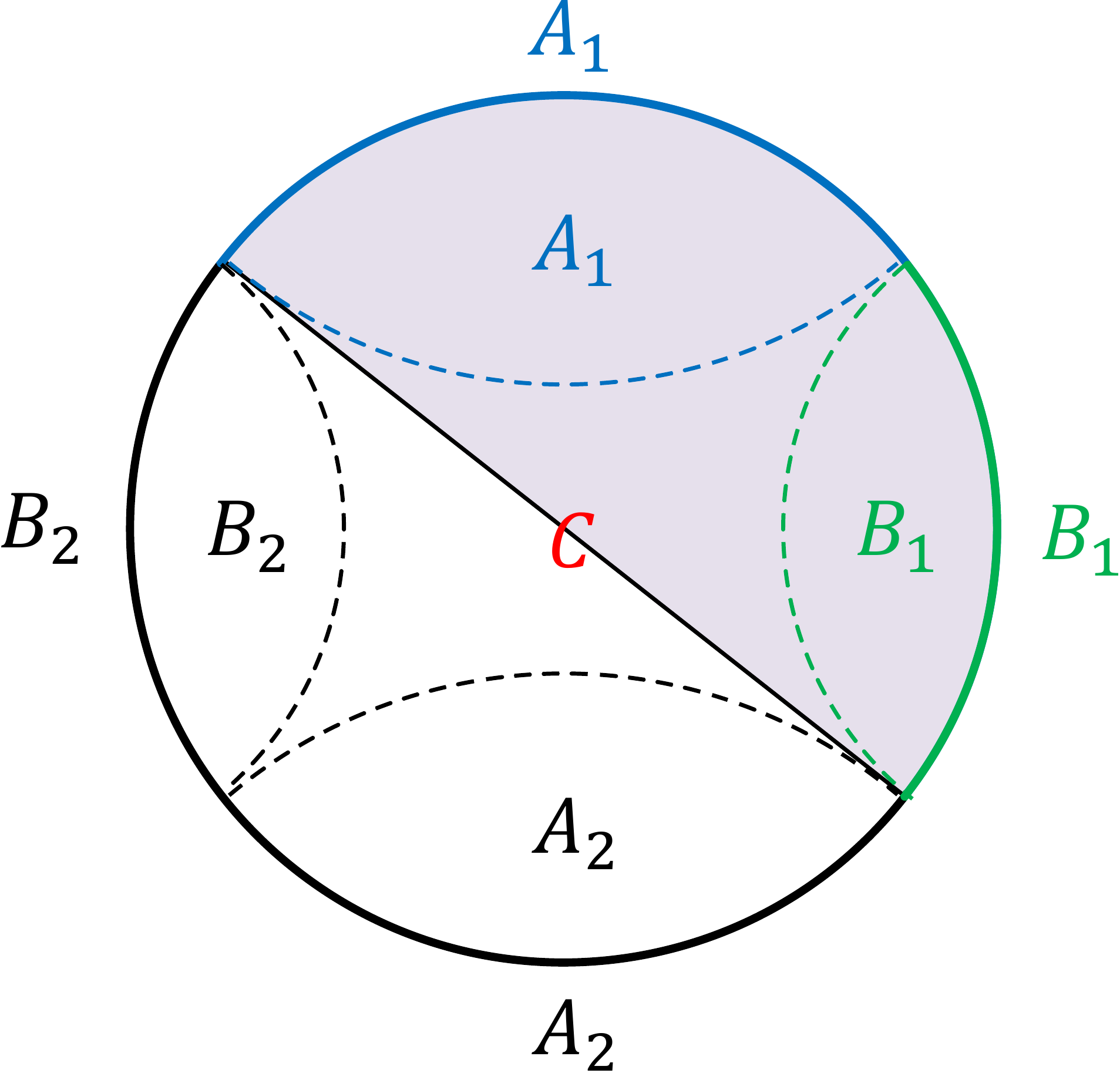}
\caption{%
The entanglement wedge of $A_1$ (left) and that of $A_1\cup B_1$ (right). 
The entanglement wedge of $A_1$ does not includes $C$, 
and similarly, the regions of each of $R_+$, $R_-$, $B_+$ or $B_-$ 
does not include the island $I$ (See Fig.~\ref{fig:BH-r}). 
The entanglement wedge of $A_1\cup B_1$ includes a half of the region $C$. 
This is similar to the region of $R_+\cup B_+$ and that of $R_+\cup B_-$ --- 
the effective region includes a half of the island in both cases. 
}\label{fig:holo2}
\end{center}
\end{figure}

However, the region $C$ exists even if the degrees of freedom of $B$ is larger than $A$. 
This is different from the island, which appears in the calculation 
of the entanglement entropy of the Hawking radiation only after the Page time. 
In order to reproduce this behavior, 
we divide the system of the black hole as in Fig.~\ref{fig:holo}(center). 
Then, the region $C$ is included in the entanglement wedge of $A$ 
only when $A$ is larger than $B=B_1\cup B_2$. 
If island appear by a similar mechanism, 
this analogy implies that the region of the black hole subsystem 
also includes a region which does not appear 
if we further divide the black hole subsystem to smaller subsystems. 
In fact, we found that the region $I$ does not appear in the entanglement entropy of 
the black hole subsystem in either right or left wedge. 
Hence, the island appears even before the Page time 
but is associated to the black hole $B = B_+ \cup B_-$. 
The region of $R_\pm \cup B_\pm$ and that of $R_\pm \cup B_\mp$ 
has also analogous structure to the holographic entanglement wedges (See Fig.~\ref{fig:holo2}).


\section{Islands in evaporating black holes}\label{sec:position}

In this section, we study the location of the island in the evaporating black holes. 
It was found in \cite{Penington:2019npb, Gautason:2020tmk, Hartman:2020swn} 
that the quantum extremal surface is located 
inside the horizon of the evaporating black holes, while 
those in the eternal black hole geometries are placed outside the horizon 
\cite{Almheiri:2019yqk, Almheiri:2019qdq, Gautason:2020tmk, Anegawa:2020ezn, Hashimoto:2020cas}. 
The difference of the position of the quantum extremal surface comes from the different vacua. 
The Hartle-Hawking vacuum is suitable for static two-sided black holes 
while the Unruh vacuum is more appropriate for evaporating black holes. 
In this section, we describe the details of calculation in the Unruh vacuum 
to see that the island is inside the horizon 
as long as the region $R$ is sufficiently far away from the horizon, 
but is extending outside the horizon if $R$ is very close to the horizon. 

For simplicity, we focus on the s-wave approximation. 
The result can be interpreted as that of a toy model in which 
the matter fields are given by two dimensional massless fields. 
It is also expected that the result would qualitatively be the same to 
that without the s-wave approximation.%
\footnote{%
The s-wave approximation is valid if the region $R$ and the island $I$ 
are sufficiently far away from each other. 
If the island is inside the horizon whereas the region $R$ is outside the horizon, 
the approximation would be valid, and hence, the island $I$ would be inside the horizon 
even without the approximation. 
If the island $I$ is outside the horizon, the distance between $R$ and $I$ would be small, 
and the s-wave approximation would be invalid. 
In this case, the other modes than s-waves gives additional attractive effects 
between the twist operators at $a_+$ and $b_+$, and then, the distance becomes smaller. 
If the island extends outside the horizon in the s-wave approximation, 
the same would be true even without the s-wave approximation. 
}


\subsection{Green function and vacuum state}

By using the replica trick, the entanglement entropy is given in terms of 
the correlation function of the twist operators. 
It can be read off from the correlation functions how 
the entanglement entropy depends on the quantum state. 
Here, we study the correlation functions in vacua. 

We consider a massless free scalar field $\phi$ in a two dimensional spacetime. 
The two point correlation function is given as 
\begin{equation}
 \left\langle \phi(x) \phi(x') \right\rangle 
 = \frac{1}{4\pi}\log \left|(u-u')(v-v')\right| \ , 
 \label{Green}
\end{equation}
up to some regular terms, where $u$ and $v$ are retarded and advanced time coordinates. 
The coordinates $(u,v)$ are not unique, but the expression above 
is exact if they are associated to the vacuum state. 

The scalar field $\phi$ is expanded in the Fourier modes in the coordinates $(u,v)$ as 
\begin{equation}
 \phi(x) 
 = 
 \int \frac{ d \omega}{2\pi} \frac{1}{\sqrt{2\omega}} 
 \left[a_\omega e^{-i\omega v} + a_\omega^\dag e^{i\omega v}
 + b_\omega e^{-i\omega u} + b_\omega^\dag e^{i\omega u}\right]
\end{equation}
where $a_\omega$ and $a_\omega^\dag$ are annihilation and creation operators 
of the infalling modes and $b_\omega$ and $b_\omega^\dag$ are those of outgoing modes. 
The vacuum state $|0\rangle$ is defined as the state which is annihilated by the annihilation operators, 
\begin{align}
 a_\omega |0\rangle &= 0 \ , 
 &
 b_\omega |0\rangle &= 0 \ . 
\end{align}
The correlation function in this vacuum state is exactly given by 
\begin{align}
 \langle 0 | \phi(x) \phi(x') |0\rangle 
 &= 
 \int \frac{ d \omega}{4\pi \omega} \left[e^{-i\omega(v-v')} + e^{-i\omega(u-u')}\right]
 \notag\\
 &= \frac{1}{4\pi}\log \left|(u-u')(v-v')\right| \ . 
\end{align}

The definition of the vacuum state depends on the coordinates 
which is used to define the annihilation and creation operators. 
Another vacuum state $|\Omega\rangle$ can be defined by using another pair of coordinates $(U,V)$, 
and then, the Green function in $|\Omega\rangle$ is different from that in $|0\rangle$; 
\begin{align}
 \langle \Omega | \phi(x) \phi(x') |\Omega \rangle 
 &= \frac{1}{4\pi}\log \left|(U-U')(V-V')\right| 
 \neq
 \langle 0 | \phi(x) \phi(x') |0\rangle \ . 
\end{align}
The difference of the correlation functions in different vacua 
is regular, and the divergent part is universally given by \eqref{Green}. 

Note that physics is independent of choice of coordinates. 
Although the vacuum state is defined by using a pair of the coordinates, 
it is independent from the coordinate in the sense that 
it is invariant under the coordinate transformation. 
Once the vacuum state $|0\rangle$ is defined in terms of $u$ and $v$, 
it is annihilated by the same annihilation operators even after the coordinate transformation. 
Thus, the Green function is given by \eqref{Green} exactly by using appropriate coordinates, 
but it depends on the vacuum state which coordinates are appropriate.


\subsection{Islands in Unruh vacuum}

Now, we consider the island in the evaporating black hole. 
In this case, the vacuum state should be the Unruh vacuum. 
It is annihilated by the annihilation operators 
which are defined by the plane waves in the flat spacetime 
before the gravitational collapse. 
The retarded time $U$ is approximately the same to that in the Kruskal coordinates, 
but the advanced time $v$ is in the advanced time in the ingoing Eddington-Finkelstein coordinates. 
The metric is approximately given in the Eddington-Finkelstein coordinate as 
\begin{equation}
 ds^2 = - \left(1-\frac{r_h(v)}{r}\right) dv^2 + 2 dv dr + r^2 d \Omega^2 \ . 
\end{equation}
The radius on outgoing null lines are give by solutions of 
\begin{equation}
 \frac{dr}{dv} = \frac{1}{2} \left(1 - \frac{r_h(v)}{r}\right) \ , 
\end{equation}
which is approximately solved near the horizon as 
\begin{equation}
 r \simeq r_h(v) - U e^{\frac{v}{2r_h(v)}} + 2 r_h(v) \frac{d r_h(v)}{dv} \ , 
\end{equation}
where $U$ is the integration constant, which is identified with the retarded time. 
The coordinate $U$ is identical to the advanced time in the flat spacetime 
before the gravitational collapse up to a constant factor 
if the black hole is formed by a thin null shell. 
It is also approximately the same to that in the Kruskal coordinate near the horizon. 
In terms of $U$ and $v$, the metric is expressed as 
\begin{align}
 ds^2 
 \simeq - 2 e^{ \frac{v}{2 r_h}} dU dv  + r^2 d \Omega^2 \ . 
\end{align}

The matter part of the entanglement entropy in the s-wave approximation is given by \eqref{S-long}, 
but now, the coordinate $V$ and the factor $W$ are replaced by 
$v$ and $e^{-\frac{v}{4r_h}}$, respectively. 
Then, $S_\text{matter}$ is obtained as 
\begin{align}
 S_\text{matter} 
 &= 
 \frac{c}{6} \log \left|(U_a-U_b)(v_a-v_b)\right| 
 + \frac{c\,v_a}{24r_h(v_a)} + \frac{c\,v_b}{24r_h(v_b)} , 
 \label{Smatter-Unruh}
\end{align}
where $U_b$ and $v_b$ stand for the position of the quantum extremal surface $a_+$, 
and $U_b$ and $v_b$ are those of $b_+$, the inner boundary of $R$. 
In \cite{Penington:2019npb} only outgoing modes are taken into the calculation. 
It is equivalent to taking the region $R$ to the future infinity, $v_b \to \infty$. 
In this limit, the expression above is consistent with that in \cite{Penington:2019npb} 
up to divergent terms which can be treated as constant terms. 

The position of the quantum extremal surface, $a_+$, is determined such that 
the total entanglement entropy becomes the extremum. 
Assuming that the time evolution of $r_h$ is given by the standard formula of the Hawking radiation, 
\begin{equation}
 \frac{d r_h}{d v} = - \frac{c\,G_N}{96\pi r_h^2} \ , 
\end{equation}
the position of the quantum extremal surface, $(U,v)$, is determined by the following two conditions; 
\begin{align}
 r_h - b &= \frac{c\,G_N}{\pi} e^{-\frac{v_a-v_b}{2r_h}} \left(-\frac{1}{16r_h} + \frac{1}{6(v_a-v_b)}\right) \ , 
\label{cond-u0-Unruh}
 \\
 U_a &= U_b \frac{8 r_h + (v_a-v_b)}{8 r_h - 3 (v_a-v_b)} \ , 
\label{cond-u-Unruh}
\end{align}
where we used 
\begin{equation}
 b \simeq r_h - U_b e^{ \frac{v_b}{2 r_h}} \ . 
\end{equation}
By using \eqref{cond-u0-Unruh}, we first calculate $v_a$, for given $(U_b,v_b)$, 
and then, $U_a$ is determined by \eqref{cond-u-Unruh}. 
Here $U_b>0$ and $v_a-v_b < 0$, 
since $b_+$ is outside the horizon and $a_+$ and $b_+$ has a spacelike separation. 
Eq.~\eqref{cond-u-Unruh} indicates that the quantum extremal surface is located 
inside the event horizon for $v_a-v_b < - 8 r_h$, but is outside the horizon for $v_a-v_b > -8r_h$. 
Eq.~\eqref{cond-u0-Unruh} gives $b - r_h = \mathcal O(G_N)$ for $v_a-v_b = - 8 r_h$, 
and then, $v_a-v_b \ll - 8 r_h$ for $b - r_h = \mathcal O(r_h)$. 
This implies that the island is located inside the horizon 
as long as the distance between the region $R$ and the horizon 
is much larger than the Planck length  
but the island extends outside the horizon if $R$ extends sufficiently close to the horizon. 

It is straightforward to see that \eqref{cond-u0-Unruh} and \eqref{cond-u-Unruh} 
have a solution with real $U_a$ and $v_a$ only for 
\begin{equation}
 b \geq b_c \ , 
\end{equation}
where 
\begin{equation}
 b_c = r_h + \frac{3 c\,G_N}{16\pi r_h} e^{2/3} \ , 
 \label{bc-unruh}
\end{equation}
implying that the quantum extremal surface becomes unstable for $b\leq b_c$.%
\footnote{
The position of the stretched horizon \eqref{bc-unruh} is much closer 
to the horizon than \eqref{bc}. 
This is because of the s-wave approximation. 
It is straightforward to repeat the same calculation to Sec.~\ref{sec:stretch} 
by using the s-wave approximation, and we obtain 
\begin{equation}
 b_c = r_h + \frac{c\,G_N}{3\pi r_h} \ . 
\end{equation}
Thus, the distance from the horizon is of the same order to that in the Hartle-Hawking vacuum. 
Without the s-wave approximation, the attraction between the twist operators becomes much stronger, 
and then, the distance is expected to be of the same order to \eqref{bc}. 
}
The unstable saddle point of the quantum extremal surface for $b=b_c$ (Fig.~\ref{fig:critical}(right)) 
is given by 
\begin{align}
 v_a &= v_b - \frac{4}{3} r_h \ , 
 & 
 U_a = \frac{5}{9} U_b \ , 
\end{align}
and hence is located outside the horizon, namely $U_a < 0$ since $U_b < 0$. 
Although the instability for $b=b_c$ implies that the region $R$ and the island $I$ becomes continuous, 
the island $I$ extends outside the horizon but disconnected from the region $R$ for $b\gtrsim b_c$. 

%

As we have seen in this section, the position of the quantum extremal surface depends on the vacuum state. 
Although the island is extending outside the horizon in the Hartle-Hawking vacuum, 
it is located inside the horizon in the Unruh vacuum as long as 
$R$ is sufficiently away from the horizon. 
However, the quantum extremal surface is placed outside the horizon 
independent of the vacuum state, if $R$ is sufficiently close to the horizon. 
This is because the position of the quantum extremal surface is 
related to the correlation function of the twist operators. 
If the twist operators are sufficiently close to each other, 
the singular part becomes most dominant and 
the correlation function becomes almost independent of the vacuum state. 
Thus the position of the quantum extremal surface in the Unruh vacuum 
is similar to that in the Hartle-Hawking vacuum, 
when $R$ is sufficiently close to the quantum extremal surface. 
This is the reason why the stretched horizon $b_c$ is always outside the horizon. 


\end{appendix}



\begin{thebibliography}{99}


\bibitem{Hawking:1976ra}
S.~Hawking,
``Breakdown of Predictability in Gravitational Collapse,''
Phys.\ Rev.\ D \textbf{14} (1976), 2460-2473.


\bibitem{Hawking:1974sw} 
  S.~W.~Hawking,
  ``Particle Creation by Black Holes,''
  Commun.\ Math.\ Phys.\  {\bf 43}, 199 (1975)
  Erratum: [Commun.\ Math.\ Phys.\  {\bf 46}, 206 (1976)].



\bibitem{Penington:2019npb}
G.~Penington,
``Entanglement Wedge Reconstruction and the Information Paradox,''
JHEP \textbf{09} (2020), 002
[arXiv:1905.08255 [hep-th]].

\bibitem{Almheiri:2019psf}
A.~Almheiri, N.~Engelhardt, D.~Marolf and H.~Maxfield,
``The entropy of bulk quantum fields and the entanglement wedge of an evaporating black hole,''
JHEP \textbf{12} (2019), 063
[arXiv:1905.08762 [hep-th]].



\bibitem{Almheiri:2019hni}
A.~Almheiri, R.~Mahajan, J.~Maldacena and Y.~Zhao,
``The Page curve of Hawking radiation from semiclassical geometry,''
JHEP \textbf{03} (2020), 149
[arXiv:1908.10996 [hep-th]].

\bibitem{Almheiri:2019yqk} 
  A.~Almheiri, R.~Mahajan and J.~Maldacena,
  ``Islands outside the horizon,''
  arXiv:1910.11077 [hep-th].



\bibitem{Penington:2019kki} 
  G.~Penington, S.~H.~Shenker, D.~Stanford and Z.~Yang,
  ``Replica wormholes and the black hole interior,''
  arXiv:1911.11977 [hep-th].

\bibitem{Almheiri:2019qdq}
A.~Almheiri, T.~Hartman, J.~Maldacena, E.~Shaghoulian and A.~Tajdini,
``Replica Wormholes and the Entropy of Hawking Radiation,''
JHEP \textbf{05} (2020), 013
[arXiv:1911.12333 [hep-th]].



\bibitem{Almheiri:2020cfm}
A.~Almheiri, T.~Hartman, J.~Maldacena, E.~Shaghoulian and A.~Tajdini,
``The entropy of Hawking radiation,''
[arXiv:2006.06872 [hep-th]].



\bibitem{Akers:2019nfi}
C.~Akers, N.~Engelhardt and D.~Harlow,
JHEP \textbf{08} (2020), 032
[arXiv:1910.00972 [hep-th]].

\bibitem{Chen:2019uhq} 
  H.~Z.~Chen, Z.~Fisher, J.~Hernandez, R.~C.~Myers and S.~M.~Ruan,
  ``Information Flow in Black Hole Evaporation,''
  JHEP {\bf 2003}, 152 (2020)
  [arXiv:1911.03402 [hep-th]].

\bibitem{Almheiri:2019psy}
A.~Almheiri, R.~Mahajan and J.~E.~Santos,
``Entanglement islands in higher dimensions,''
SciPost Phys. \textbf{9} (2020) no.1, 001
[arXiv:1911.09666 [hep-th]].
  
\bibitem{Chen:2019iro} 
  Y.~Chen,
  ``Pulling Out the Island with Modular Flow,''
  JHEP {\bf 2003}, 033 (2020)
  [arXiv:1912.02210 [hep-th]].

\bibitem{Akers:2019lzs}
C.~Akers, N.~Engelhardt, G.~Penington and M.~Usatyuk,
``Quantum Maximin Surfaces,''
JHEP \textbf{08} (2020), 140
[arXiv:1912.02799 [hep-th]].

\bibitem{Liu:2020gnp}
H.~Liu and S.~Vardhan,
``A dynamical mechanism for the Page curve from quantum chaos,''
[arXiv:2002.05734 [hep-th]].
  
\bibitem{Marolf:2020xie}
D.~Marolf and H.~Maxfield,
``Transcending the ensemble: baby universes, spacetime wormholes, and the order and disorder of black hole information,''
JHEP \textbf{08} (2020), 044
[arXiv:2002.08950 [hep-th]].

\bibitem{Balasubramanian:2020hfs}
  V.~Balasubramanian, A.~Kar, O.~Parrikar, G.~S\'arosi and T.~Ugajin,
  ``Geometric secret sharing in a model of Hawking radiation,''
  arXiv:2003.05448 [hep-th].

\bibitem{Bhattacharya:2020ymw}
A.~Bhattacharya,
``Multipartite purification, multiboundary wormholes, and islands in $AdS_3/CFT_2$,''
Phys. Rev. D \textbf{102} (2020) no.4, 046013
[arXiv:2003.11870 [hep-th]].
  
\bibitem{Verlinde:2020upt} 
  H.~Verlinde,
  ``ER = EPR revisited: On the Entropy of an Einstein-Rosen Bridge,''
  arXiv:2003.13117 [hep-th].

\bibitem{Chen:2020wiq}
Y.~Chen, X.~L.~Qi and P.~Zhang,
``Replica wormhole and information retrieval in the SYK model coupled to Majorana chains,''
JHEP \textbf{06} (2020), 121
[arXiv:2003.13147 [hep-th]].
  
\bibitem{Gautason:2020tmk}
F.~F.~Gautason, L.~Schneiderbauer, W.~Sybesma and L.~Thorlacius,
``Page Curve for an Evaporating Black Hole,''
JHEP \textbf{05} (2020), 091
[arXiv:2004.00598 [hep-th]].

\bibitem{Anegawa:2020ezn}
T.~Anegawa and N.~Iizuka,
``Notes on islands in asymptotically flat 2d dilaton black holes,''
JHEP \textbf{07} (2020), 036
[arXiv:2004.01601 [hep-th]].

\bibitem{Hashimoto:2020cas}
K.~Hashimoto, N.~Iizuka and Y.~Matsuo,
``Islands in Schwarzschild black holes,''
JHEP \textbf{06} (2020), 085
[arXiv:2004.05863 [hep-th]].

\bibitem{Sully:2020pza}
J.~Sully, M.~Van Raamsdonk and D.~Wakeham,
``BCFT entanglement entropy at large central charge and the black hole interior,''
[arXiv:2004.13088 [hep-th]].

\bibitem{Hartman:2020swn}
T.~Hartman, E.~Shaghoulian and A.~Strominger,
``Islands in Asymptotically Flat 2D Gravity,''
JHEP \textbf{07} (2020), 022
[arXiv:2004.13857 [hep-th]].

\bibitem{Hollowood:2020cou}
T.~J.~Hollowood and S.~P.~Kumar,
``Islands and Page Curves for Evaporating Black Holes in JT Gravity,''
JHEP \textbf{08} (2020), 094
[arXiv:2004.14944 [hep-th]].

\bibitem{Krishnan:2020oun}
C.~Krishnan, V.~Patil and J.~Pereira,
[arXiv:2005.02993 [hep-th]].

\bibitem{Alishahiha:2020qza}
M.~Alishahiha, A.~Faraji Astaneh and A.~Naseh,
``Island in the Presence of Higher Derivative Terms,''
[arXiv:2005.08715 [hep-th]].

\bibitem{Banks:2020zrt}
T.~Banks,
``Microscopic Models of Linear Dilaton Gravity and Their Semi-classical Approximations,''
[arXiv:2005.09479 [hep-th]].

\bibitem{Geng:2020qvw}
H.~Geng and A.~Karch,
``Massive islands,''
JHEP \textbf{09} (2020), 121
[arXiv:2006.02438 [hep-th]].

\bibitem{Chen:2020uac}
H.~Z.~Chen, R.~C.~Myers, D.~Neuenfeld, I.~A.~Reyes and J.~Sandor,
``Quantum Extremal Islands Made Easy, Part I: Entanglement on the Brane,''
JHEP \textbf{10} (2020), 166
[arXiv:2006.04851 [hep-th]].

\bibitem{Chandrasekaran:2020qtn}
V.~Chandrasekaran, M.~Miyaji and P.~Rath,
``Including contributions from entanglement islands to the reflected entropy,''
Phys. Rev. D \textbf{102} (2020) no.8, 086009
[arXiv:2006.10754 [hep-th]].

\bibitem{Li:2020ceg}
T.~Li, J.~Chu and Y.~Zhou,
``Reflected Entropy for an Evaporating Black Hole,''
[arXiv:2006.10846 [hep-th]].

\bibitem{Bak:2020enw}
D.~Bak, C.~Kim, S.~H.~Yi and J.~Yoon,
``Unitarity of Entanglement and Islands in Two-Sided Janus Black Holes,''
[arXiv:2006.11717 [hep-th]].

\bibitem{Bousso:2020kmy}
R.~Bousso and E.~Wildenhain,
``Gravity/ensemble duality,''
Phys. Rev. D \textbf{102} (2020) no.6, 066005
[arXiv:2006.16289 [hep-th]].

\bibitem{Anous:2020lka}
T.~Anous, J.~Kruthoff and R.~Mahajan,
``Density matrices in quantum gravity,''
SciPost Phys. \textbf{9} (2020) no.4, 045
[arXiv:2006.17000 [hep-th]].

\bibitem{Dong:2020uxp}
X.~Dong, X.~L.~Qi, Z.~Shangnan and Z.~Yang,
``Effective entropy of quantum fields coupled with gravity,''
JHEP \textbf{10} (2020), 052
[arXiv:2007.02987 [hep-th]].

\bibitem{Hollowood:2020kvk}
T.~J.~Hollowood, S.~Prem Kumar and A.~Legramandi,
``Hawking Radiation Correlations of Evaporating Black Holes in JT Gravity,''
J. Phys. A \textbf{53} (2020) no.47, 475401
[arXiv:2007.04877 [hep-th]].

\bibitem{Krishnan:2020fer}
C.~Krishnan,
[arXiv:2007.06551 [hep-th]].
\bibitem{Engelhardt:2020qpv}
N.~Engelhardt, S.~Fischetti and A.~Maloney,
``Free Energy from Replica Wormholes,''
[arXiv:2007.07444 [hep-th]].

\bibitem{Karlsson:2020uga}
A.~Karlsson,
``Replica wormhole and island incompatibility with monogamy of entanglement,''
[arXiv:2007.10523 [hep-th]].

\bibitem{Chen:2020jvn}
H.~Z.~Chen, Z.~Fisher, J.~Hernandez, R.~C.~Myers and S.~M.~Ruan,
``Evaporating Black Holes Coupled to a Thermal Bath,''
[arXiv:2007.11658 [hep-th]].

\bibitem{Chen:2020tes}
Y.~Chen, V.~Gorbenko and J.~Maldacena,
``Bra-ket wormholes in gravitationally prepared states,''
[arXiv:2007.16091 [hep-th]].

\bibitem{Hartman:2020khs}
T.~Hartman, Y.~Jiang and E.~Shaghoulian,
``Islands in cosmology,''
[arXiv:2008.01022 [hep-th]].

\bibitem{Liu:2020jsv}
H.~Liu and S.~Vardhan,
``Entanglement entropies of equilibrated pure states in quantum many-body systems and gravity,''
[arXiv:2008.01089 [hep-th]].

\bibitem{Murdia:2020iac}
C.~Murdia, Y.~Nomura and P.~Rath,
``Coarse-Graining Holographic States: A Semiclassical Flow in General Spacetimes,''
Phys. Rev. D \textbf{102} (2020) no.8, 086001
[arXiv:2008.01755 [hep-th]].

\bibitem{Akers:2020pmf}
C.~Akers and G.~Penington,
``Leading order corrections to the quantum extremal surface prescription,''
[arXiv:2008.03319 [hep-th]].

\bibitem{Balasubramanian:2020xqf}
V.~Balasubramanian, A.~Kar and T.~Ugajin,
``Islands in de Sitter space,''
[arXiv:2008.05275 [hep-th]].

\bibitem{Balasubramanian:2020coy}
V.~Balasubramanian, A.~Kar and T.~Ugajin,
``Entanglement between two disjoint universes,''
[arXiv:2008.05274 [hep-th]].

\bibitem{Sybesma:2020fxg}
W.~Sybesma,
``Pure de Sitter space and the island moving back in time,''
[arXiv:2008.07994 [hep-th]].

\bibitem{Stanford:2020wkf}
D.~Stanford,
``More quantum noise from wormholes,''
[arXiv:2008.08570 [hep-th]].

\bibitem{Chen:2020hmv}
H.~Z.~Chen, R.~C.~Myers, D.~Neuenfeld, I.~A.~Reyes and J.~Sandor,
``Quantum Extremal Islands Made Easy, Part II: Black Holes on the Brane,''
[arXiv:2010.00018 [hep-th]].

\bibitem{Ling:2020laa}
Y.~Ling, Y.~Liu and Z.~Y.~Xian,
``Island in Charged Black Holes,''
[arXiv:2010.00037 [hep-th]].

\bibitem{Marolf:2020rpm}
D.~Marolf and H.~Maxfield,
``Observations of Hawking radiation: the Page curve and baby universes,''
[arXiv:2010.06602 [hep-th]].

\bibitem{Harlow:2020bee}
D.~Harlow and E.~Shaghoulian,
``Global symmetry, Euclidean gravity, and the black hole information problem,''
[arXiv:2010.10539 [hep-th]].

\bibitem{Akal:2020ujg}
I.~Akal,
``Universality, intertwiners and black hole information,''
[arXiv:2010.12565 [hep-th]].

\bibitem{Hernandez:2020nem}
J.~Hernandez, R.~C.~Myers and S.~M.~Ruan,
``Quantum Extremal Islands Made Easy, PartIII: Complexity on the Brane,''
[arXiv:2010.16398 [hep-th]].



\bibitem{Callan:1994py}
C.~G.~Callan, Jr. and F.~Wilczek,
``On geometric entropy,''
Phys.\ Lett.\ B \textbf{333} (1994), 55-61
[arXiv:hep-th/9401072 [hep-th]].

\bibitem{Holzhey:1994we}
C.~Holzhey, F.~Larsen and F.~Wilczek,
``Geometric and renormalized entropy in conformal field theory,''
Nucl.\ Phys.\ B \textbf{424} (1994), 443-467
[arXiv:hep-th/9403108 [hep-th]].

\bibitem{Calabrese:2009qy}
P.~Calabrese and J.~Cardy,
``Entanglement entropy and conformal field theory,''
J.\ Phys.\ A \textbf{42} (2009), 504005
[arXiv:0905.4013 [cond-mat.stat-mech]].



\bibitem{Engelhardt:2014gca}
N.~Engelhardt and A.~C.~Wall,
``Quantum Extremal Surfaces: Holographic Entanglement Entropy beyond the Classical Regime,''
JHEP \textbf{01} (2015), 073
[arXiv:1408.3203 [hep-th]].


\bibitem{Ryu:2006bv}
S.~Ryu and T.~Takayanagi,
``Holographic derivation of entanglement entropy from AdS/CFT,''
Phys.\ Rev.\ Lett.\  \textbf{96} (2006), 181602
[arXiv:hep-th/0603001 [hep-th]].

\bibitem{Hubeny:2007xt}
V.~E.~Hubeny, M.~Rangamani and T.~Takayanagi,
``A Covariant holographic entanglement entropy proposal,''
JHEP \textbf{07} (2007), 062
[arXiv:0705.0016 [hep-th]].



\bibitem{Faulkner:2013ana}
T.~Faulkner, A.~Lewkowycz and J.~Maldacena,
``Quantum corrections to holographic entanglement entropy,''
JHEP \textbf{11} (2013), 074
[arXiv:1307.2892 [hep-th]].

\bibitem{Lewkowycz:2013nqa}
A.~Lewkowycz and J.~Maldacena,
``Generalized gravitational entropy,''
JHEP \textbf{08} (2013), 090
[arXiv:1304.4926 [hep-th]].

\bibitem{Dong:2016hjy}
X.~Dong, A.~Lewkowycz and M.~Rangamani,
``Deriving covariant holographic entanglement,''
JHEP \textbf{11} (2016), 028
[arXiv:1607.07506 [hep-th]].

\bibitem{Dong:2017xht}
X.~Dong and A.~Lewkowycz,
``Entropy, Extremality, Euclidean Variations, and the Equations of Motion,''
JHEP \textbf{01} (2018), 081
[arXiv:1705.08453 [hep-th]].



\bibitem{Bombelli:1986rw}
L.~Bombelli, R.~K.~Koul, J.~Lee and R.~D.~Sorkin,
``A Quantum Source of Entropy for Black Holes,''
Phys.\ Rev.\ D \textbf{34} (1986), 373-383.

\bibitem{Srednicki:1993im}
M.~Srednicki,
``Entropy and area,''
Phys.\ Rev.\ Lett.\  \textbf{71} (1993), 666-669
[arXiv:hep-th/9303048 [hep-th]].

  
\bibitem{Susskind:1994sm} 
  L.~Susskind and J.~Uglum,
  ``Black hole entropy in canonical quantum gravity and superstring theory,''
  Phys.\ Rev.\ D {\bf 50}, 2700 (1994)
  [hep-th/9401070].

  
  
\bibitem{Casini:2005zv} 
  H.~Casini and M.~Huerta,
  ``Entanglement and alpha entropies for a massive scalar field in two dimensions,''
  J.\ Stat.\ Mech.\  {\bf 0512}, P12012 (2005)
  [cond-mat/0511014].

\bibitem{Casini:2009sr} 
  H.~Casini and M.~Huerta,
  ``Entanglement entropy in free quantum field theory,''
  J.\ Phys.\ A {\bf 42}, 504007 (2009)
  [arXiv:0905.2562 [hep-th]].
  


\bibitem{Page:1993wv} 
  D.~N.~Page,
  ``Information in black hole radiation,''
  Phys.\ Rev.\ Lett.\  {\bf 71}, 3743 (1993)
  [hep-th/9306083].

\bibitem{Page:2013dx} 
  D.~N.~Page,
  ``Time Dependence of Hawking Radiation Entropy,''
  JCAP {\bf 1309}, 028 (2013)
  [arXiv:1301.4995 [hep-th]].



\bibitem{Czech:2012bh}
B.~Czech, J.~L.~Karczmarek, F.~Nogueira and M.~Van Raamsdonk,
``The Gravity Dual of a Density Matrix,''
Class. Quant. Grav. \textbf{29} (2012), 155009
[arXiv:1204.1330 [hep-th]].

\bibitem{Wall:2012uf}
A.~C.~Wall,
``Maximin Surfaces, and the Strong Subadditivity of the Covariant Holographic Entanglement Entropy,''
Class. Quant. Grav. \textbf{31} (2014) no.22, 225007
[arXiv:1211.3494 [hep-th]].

\bibitem{Headrick:2014cta}
M.~Headrick, V.~E.~Hubeny, A.~Lawrence and M.~Rangamani,
``Causality \& holographic entanglement entropy,''
JHEP \textbf{12} (2014), 162
[arXiv:1408.6300 [hep-th]].



\bibitem{Damour:1978cg} 
  T.~Damour,
  ``Black Hole Eddy Currents,''
  Phys.\ Rev.\ D {\bf 18}, 3598 (1978).

\bibitem{Damour:1982} 
  T.~Damour, 
  ``Surface Effects in Black Hole Physics,''
  in {\it Proceedings of the Second Marcel Grossman Meeting on General Relativity}, 
  ed. R. Ruffini (North-Holland, Amsterdam, 1982), 587.

\bibitem{Price:1986yy}
  R.~H.~Price and K.~S.~Thorne,
  ``Membrane Viewpoint On Black Holes: Properties And Evolution Of The Stretched Horizon,''
  Phys.\ Rev.\ D {\bf 33} (1986) 915.

\bibitem{Thorne:1986iy}
K.~S.~Thorne, R.~H.~Price and D.~A.~Macdonald,
``BLACK HOLES: THE MEMBRANE PARADIGM,''
Yale University Press (1986). 
  
\bibitem{Parikh:1997ma}
  M.~Parikh and F.~Wilczek,
  ``An Action for black hole membranes,''
  Phys.\ Rev.\  D {\bf 58} (1998) 064011
  [arXiv:gr-qc/9712077].



\bibitem{Susskind:1993if}
L.~Susskind, L.~Thorlacius and J.~Uglum,
``The Stretched horizon and black hole complementarity,''
Phys. Rev. D \textbf{48} (1993), 3743-3761
[arXiv:hep-th/9306069 [hep-th]].

\bibitem{Stephens:1993an}
C.~R.~Stephens, G.~'t Hooft and B.~F.~Whiting,
``Black hole evaporation without information loss,''
Class. Quant. Grav. \textbf{11} (1994), 621-648
[arXiv:gr-qc/9310006 [gr-qc]].



\bibitem{Almheiri:2012rt}
A.~Almheiri, D.~Marolf, J.~Polchinski and J.~Sully,
``Black Holes: Complementarity or Firewalls?,''
JHEP \textbf{02} (2013), 062
[arXiv:1207.3123 [hep-th]].

\end{thebibliography}
\end{document}